\documentclass[twocolumn,aps,prb]{revtex4}
\usepackage{epsfig,graphicx}
\usepackage{amsmath,verbatim}
\usepackage{dcolumn}

\begin{document}

%\title{Reactive force field for lithium-aluminum \\
%silicates with applications to eucryptite phases}

\title{\fontfamily{phv}\fontseries{b}\selectfont Reactive force field for lithium-aluminum \\
silicates with applications to eucryptite phases\normalfont}

\author{Badri Narayanan,$^1$ Adri C.~T. van Duin,$^2$ Branden B. Kappes,$^3$ Ivar E. Reimanis,$^1$ and Cristian V. Ciobanu$^{3}$\footnote{Corresponding author, email: cciobanu@mines.edu}}
\affiliation{$^1$Department of Metallurgical \& Materials Engineering, Colorado School of Mines, Golden, Colorado 80401 \\
$^2$Department of Mechanical Engineering, Pennsylvania State University, University Park, Pennsylvania 16802 \\
$^3$Division of Engineering, Colorado School of Mines, Golden, Colorado 80401}

\begin{abstract}
We have parameterized a reactive force field (ReaxFF) for lithium aluminum silicates using
density functional theory (DFT) calculations of structural properties of a number of bulk phase oxides,
silicates, and aluminates, as well as of several representative clusters. The force field parameters
optimized in this study were found to predict lattice parameters and heats of formation of selected
condensed phases in excellent agreement with previous DFT calculations and with experiments.
We have used the newly developed force-field to study the eucryptite phases in terms of their
thermodynamic stability and their elastic properties. We have found that (a) these ReaxFF parameters predict
the correct order of stability of the three crystalline polymorphs of eucryptite, $\alpha$, $\beta$, and $\gamma$,
and (b) that upon indentation, a new phase appears at applied pressures $\ge$ 7 GPa. The high-pressure phase
obtained upon indentation is amorphous, as illustrated by the radial distribution functions calculated for
different pairs of elements.
In terms of elastic properties analysis, we have determined the elements of the stiffness tensor for $\alpha$- and $\beta$-
eucryptite at the level of ReaxFF, and discussed the elastic anisotropy of these two polymorphs.
Polycrystalline average properties of these eucryptite phases are also reported to serve as ReaxFF
predictions of their elastic moduli (in the case of $\alpha$-eucryptite), or as tests against values
known from experiments or DFT calculations ($\beta$-eucrypite). The ReaxFF potential reported here can also describe
well single-species systems ({\em e.g.}, Li-metal, Al-metal, and condensed phases of silicon), which makes it
suitable for investigating structure and properties of suboxides, atomic-scale mechanisms responsible for
phase transformations, as well as oxidation-reduction reactions.
\end{abstract}
\maketitle

\section{Introduction}
Lithium Aluminum Silicate (LAS) glass ceramics have been  investigated extensively over the last
few decades owing to their exotic properties, such as small (or slightly negative) coefficient
of thermal expansion and exceptional thermal stability.~\cite{Bach,Palmer,Xu1999}
Such unique physical properties make LAS ceramics suitable for a variety of applications such
as heat exchangers with high thermal shock resistance, high precision optical devices,
telescope mirror blanks, and ring laser gyroscopes.~\cite{Bach,Palmer,Xu1999,Lichtenstein1998,Lichtenstein2000}
$\beta$-eucryptite (LiAlSiO$_4$) is an important LAS glass ceramic material with a
hexagonal crystal structure which can be viewed as a stuffed derivative of the
high-temperature $\beta$-quartz configuration.\cite{Winkler,Buerger,Schulz_B28_1972_II,Tscherry_161_1972,Tscherry_175_1972,Pillars1973}
The structure of $\beta$-eucryptite confers it superionic conductivity of Li$^{\mathrm{+}}$ ions along the
$c$-axis, thus making it a potential electrolyte material for Li ion batteries.\cite{Alpen1977,Schulz1980,Nagel1982,Renker1983,Sartbaeva2004}

From a more fundamental perspective, $\beta$-eucryptite is known to undergo a reversible
order-disorder transformation at $\sim$755 K which occurs via spatial disordering
of the lithium atoms at high temperatures.\cite{Schulz1980}  Recently, it was reported that
$\beta$-eucryptite undergoes a reversible pressure-induced transition to a
metastable polymorph (called the $\epsilon$ phase) at $\sim$0.8 GPa.\cite{Zhang2002, Jochum2009}
Apart from $\beta$ and the recently discovered high pressure $\epsilon$ phases, there
are other known polymorphs of eucryptite, {\em i.e. } $\alpha$ and $\gamma$. Of
these, $\alpha$ is the most stable under ambient conditions and
exists over a wide range of temperatures but is typically kinetically
hindered.\cite{Beall} On the other hand, $\gamma$-eucryptite is a metastable phase
which coexists along with $\beta$-eucryptite over a narrow range of
temperatures (1038-1103 K) and is, therefore, of lower practical significance.\cite{Zhang2005}
$\alpha$-eucryptite has a rhombohedral crystal structure belonging to the $R3$ space
group similar to phenakite and willemite; its thermodynamic, structural
and physical properties have been studied.\cite{Daniels2001,Fasshauer1998} However, its elastic properties, in particular the single-crystal elastic
constants are not yet known. Furthermore, the extent of elastic anisotropy
in $\alpha$-eucryptite has also not yet been reported. Such a study could provide insight into the
structure-property relationship in LAS glass ceramics and assist in designing composites with tailored elastic properties.

The potential development work presented here has emerged from our long term goal of determining the atomic
structure of the $\epsilon$ phase and the atomic-scale mechanism responsible for the $\beta$-to-$\epsilon$ phase transformation.
While phase transformations can be directly evidenced in DFT-based Carr-Parrinello molecular dynamics (CPMD) simulations ({\em e.g.}, Refs.~\onlinecite{m1}, \onlinecite{m2}), the number of atoms in the unit cell of $\beta$ eucryptite makes it impractical to undertake such simulations in which usually several unit cells are required along each spatial direction. We have therefore not resorted to CPMD approaches,
but turned to molecular dynamics (MD) simulations based on empirical potentials.
To describe the interatomic forces acting during MD simulations, several empirical force fields (EFFs) have been proposed for LAS systems.\cite{Lewis1985,Huang1990,Beest1990,Winkler_THB1991,Cormack1996,Cormack1999,Pedone2006}
These EFFs reproduce well short-range order, mechanical, and transport properties, but may not provide a
sufficiently adequate description of phase transformations, medium range order, and vibrational density of states.
Recently, van Duin and co-workers\cite{VanDuin2001,VanDuin2003} developed a reactive force
field (ReaxFF) based on a bond-order formalism\cite{Tersoff1988,Benner1990} in conjunction with a charge
equilibration scheme.\cite{Rappe1991} In this study, we have parameterized ReaxFF for lithium aluminum silicates by fitting
against formation energies, atomic configurations, and charge distributions of a number of representative
clusters and equations of state of well-known condensed phases of oxides, silicates, and aluminates derived from DFT calculations.

This article is organized as follows. Sec.~\ref{sec:methodology} describes the methodology we
adopted to parameterize the ReaxFF for lithium aluminum silicates, the DFT data used to construct the training set for
the parametrization of ReaxFF, and the details of these DFT calculations. The parameters are given in a format compatible with the MD package LAMMPS.\cite{LAMMPS}
Sec.~\ref{sec:results} reports our results in for: heats of formation
and geometric parameters for a number of bulk phases; relative stability of eucryptite phases and
amorphization of $\beta$-eucryptite upon indentations; and elastic properties of the two most stable eucryptite
polymorphs. We have determined the elements of the stiffness tensor for $\alpha$- and $\beta$-eucryptite at the
level of ReaxFF, and discussed the elastic anisotropy of these two polymorphs. Polycrystalline average
properties of these eucryptite phases are also reported to serve as ReaxFF predictions of their elastic moduli
(in the case of $\alpha$-eucryptite), or as tests against values
known from experiments or DFT calculations ($\beta$-eucrypite). Sec.~\ref{sec:conclusion} summarizes the results and
discusses the main successes and shortcomings of our ReaxFF parametrization for LAS systems.

\section{Methodology} \label{sec:methodology}

\subsection{ReaxFF framework}
The energy contributions in ReaxFF are functions of bond orders,
which allows for a better description of bond breaking and bond formation during simulations.
Furthermore, Coulomb interactions are computed for every atom pair based on charges calculated at
every time step using a charge equilibration scheme which allows it to describe covalent, metallic, and
ionic systems equally well.\cite{VanDuin2001,VanDuin2003} The ReaxFF framework has been applied successfully to predict
the dynamics and reactive processes in hydrocarbons,\cite{VanDuin2001,Strachan2005} crack propagation
in silicon crystals,\cite{Buehler2006} interfacial reactions in Si/Si-oxide\cite{VanDuin2003}
and Al/Al-oxide,\cite{Zhang2004} surface reactions in ZnO,\cite{Raymand2008} oxygen-ion transport
in Y-stabilized ZrO$_2$,\cite{VanDuin2008} and phase transitions in ferroelectric
BaTiO$_3$.\cite{Goddard2002} The parameters of the ReaxFF potential are optimized by
fitting against density functional theory (DFT) data for various bulk phases and atomic clusters.

The formulation of ReaxFF is based on the concept of bond order,\cite{Pauling1947} which describes
the number of electrons shared between two atoms as a continuous function of their spacing.
The bond order $BO_{ij}^{\prime}$ associated with atoms $i$ and $j$ is calculated via
\begin{eqnarray}
\label{Eq:bond_order}
{BO}^{\prime}_{ij} & = & {BO}^{\prime \sigma}_{ij} + {BO}^{\prime \pi}_{ij} + {BO}^{\prime \pi\pi}_{ij} \nonumber \\
		   & = & \mathrm{exp}\left(p_{bo1}\left(\frac{r_{ij}}{r_{0}^{\sigma}}\right)^{p_{bo2}}\right) + \mathrm{exp}\left(p_{bo3}\left(\frac{r_{ij}}{r_{0}^{\pi}}\right)^{p_{bo4}}\right) \nonumber \\
		   &   &  + \mathrm{exp}\left(p_{bo5}\left(\frac{r_{ij}}{r_{0}^{\pi\pi}}\right)^{p_{bo6}}\right),
\end{eqnarray}
where ${BO}^{\prime \sigma}_{ij}$, ${BO}^{\prime \pi}_{ij}$ and ${BO}^{\prime \pi\pi}_{ij}$ are the partial contributions of $\sigma$, $\pi$- and double $\pi$-bonds involving the atoms $i$ and $j$, $r_{ij}$ is the distance between $i$ and $j$, $r_0^\sigma$, $r_0^\pi$, $r_0^{\pi\pi}$ are the bond radii of $\sigma$, $\pi$- and double $\pi$-bonds, respectively, and $p_{bo}$ are the bond order parameters. The bond orders $BO_{ij}^{\prime}$ obtained from Eq.~(\ref{Eq:bond_order}) are corrected to account for local overcoordination and residual 1--3 interactions by employing a scheme detailed in Ref.~\onlinecite{vanDuin_book}.

The total energy $E$ of the system is expressed as the sum of partial energy contributions corresponding to bonded and unbonded interactions:\cite{VanDuin2001,VanDuin2003,Strachan2005}
\begin{eqnarray}
\label{eq:total_energy}
E & = & \displaystyle\sum_{\substack{i,j\\ i<j}}E_{b,ij} + \displaystyle\sum_i E_{ov,i} + \displaystyle\sum_i E_{un,i} + \displaystyle\sum_i E_{lp,i}\nonumber \\			
                    &   &  + \displaystyle\sum_{\substack{i,j,k \\ i<j<k}}E_{v,ijk} + \displaystyle\sum_{\substack{i,j\\ i<j}}E_{vdW,ij} + \displaystyle\sum_{\substack{i,j\\ i<j}}E_{C,ij},
\end{eqnarray}
where $E_{b,ij}$ is the energy of the $i$-$j$ bond, $E_{ov,i}$ and $E_{un,i}$ are penalties for over- and under- coordination of atom $i$, $E_{lp,i}$ is the energy associated with lone-pair electrons around an atom $i$, $E_{v,ijk}$ is the energy associated with the deviation of the angle subtended at $j$ by atoms $i$ and $k$ from its equilibrium value, $E_{vdW,ij}$ and $E_{C,ij}$ are the contributions from van der Waals and Coulomb interactions between $i$ and $j$.

The energy of the $i$-$j$ bond is calculated using the corrected bond orders $BO_{ij}$ as
\begin{eqnarray}
\label{eq:bond_energy}
E_{b,ij} & = & -D^{\sigma}_{e}BO_{ij}^{\sigma}\mathrm{exp}\left(p_{be1}\left(1 - (BO_{ij}^{\sigma})^{p_{be2}}\right)\right) \nonumber \\
	 &   & -D_{e}^{\pi}BO_{ij}^{\pi} - D_{e}^{\pi\pi}BO^{\pi\pi}_{ij},
\end{eqnarray}
where $D^{\sigma}_{e}$, $D_{e}^{\pi}$ and $D_{e}^{\pi\pi}$ are the dissociation energies of $\sigma$, $\pi$- and double $\pi$-bonds, while $p_{be1,2}$ are the bond energy parameters. The contribution associated with lone pair electrons is calculated as:
\begin{equation}
E_{lp,i} = \frac{p_{lp2}\left(n_{lp,opt} - n_{lp,i}\right)}{1 + \mathrm{exp}\left(-75\left(n_{lp,opt}-n_{lp,i}\right)\right)},
\end{equation}
where $n_{lp,opt}$ is the optimum number of lone pairs for a given atom $i$ and $n_{lp,i}$ is the number of lone pairs around $i$ calculated using the relation $n_{lp,i} = \left\lfloor\frac{\Delta_i^e}{2}\right\rfloor + \mathrm{exp}\left(-p_{lp1}\left(2 + \Delta_i^e - 2\,\left\lfloor\frac{\Delta_i^e}{2}\right\rfloor\right)^2\right)$ where $\Delta_i^e$ is the difference between the number of outer shell electrons and the sum of bond orders around atom $i$ and $\lfloor x \rfloor$ is the greatest integer smaller than $x$. The penalty terms for overcoordination ($E_{ov,i}$) and undercoordination ($E_{un,i}$) of atom $i$ can be written as
\begin{subequations}
\label{eq:over_under}
\begin{equation}
E_{ov,i} = \frac{\Delta_i^{lpc} \displaystyle\sum_{j=1}^{nbond}p_{1}D^\sigma_eBO_{ij}}{\left( \Delta_i^{lpc} + \mathcal{V}_i \right)\left(1+\mathrm{exp}\left(p_{2}\Delta_i^{lpc}\right)\right)}
\end{equation}
\begin{equation}
E_{un,i} = \frac{-p_5F_{un1}(\Delta_i^{lpc})}{1 + p_7\mathrm{exp}\left(p_8F_{un2}(BO_{ij})\right)}
\end{equation}
\begin{equation}
F_{un1}(\Delta_i^{lpc}) \equiv \frac{1-\mathrm{exp}\left(p_{6}\Delta_i^{lpc}\right)}{1+\mathrm{exp}\left(-p_{2}\Delta_i^{lpc}\right)}
\end{equation}
\begin{equation}
F_{un2}(BO_{ij}) \equiv \displaystyle\sum_{j=1}^{ngb(i)}\left (\Delta_j - \Delta_j^{lp}\right) \left(BO_{ij}^{\pi}+BO_{ij}^{\pi\pi}\right)
\end{equation}
\end{subequations}
where $\Delta_j^{lp} = n_{lp,opt} - n_{lp,j}$, $\mathcal{V}_i$ is the valence of atom $i$, $\Delta_i$ is the degree of overcoordination around the atom $i$ which is corrected for the effect of broken electron pairs to obtain $\Delta_i^{lpc}$, and $p$'s are over/under coordination parameters. The energy contribution from the valence angles is written as:
\begin{subequations}
\label{eq:valence}
\begin{equation}
E_{v,ijk} = f_7\left(BO_{ij}\right)f_7\left(BO_{jk}\right)f_8\left(\Delta_j\right)F_v\left(\Theta_{ijk}\right)
\end{equation}
\begin{equation}
F_v\left(\Theta_{ijk}\right) = p_{v1}\left\{1 - \mathrm{exp}\left(-p_{v2}\left(\Theta_0 - \Theta_{ijk}\right)^2\right)\right\}
\end{equation}
\end{subequations}
where $\Theta_{ijk}$ is the angle subtended at central atom $j$ by the atoms $i$ and $k$, $\Theta_0$ is the equilibrium value for $\Theta_{ijk}$ which depends on the sum of $\pi$-bond orders (\emph{i.e.}, $BO^\pi$ and $BO^{\pi\pi}$) around the atom $j$, $f_7$ and $f_8$ are functions of bond order and degree of overcoordination, respectively, and $p_v$'s are valence angle parameters.

All the terms on the right side of Eq.~(\ref{eq:total_energy}) except the van der Waals and Coulomb interactions depend on bond order through Eqs.~(\ref{eq:bond_energy})--(\ref{eq:valence}). The bond orders are updated after every time step in a molecular dynamics simulation; such a formalism allows for a realistic simulation of dissociation and formation of bonds during a chemical reaction and also provides a good description of the bulk phases.\cite{VanDuin2001,VanDuin2003,Strachan2005}  The pairwise non-bonded interaction terms, {\em i.e.}, Coulomb and van der Waals interactions are evaluated for every atom pair irrespective of the geometry and instantaneous connectivity. The van der Waals interaction of atoms $i$ and $j$ is evaluated as

\begin{eqnarray}
E_{vdW,ij} & = & \mathcal{T}(r_{ij})D_{ij}\left\{\mathrm{exp}\left(\alpha_{ij}\left(1-\frac{f_{13}(r_{ij})}{r_{vdW}}\right)\right)\right. \nonumber \\
	   &   & \left. -2\,\mathrm{exp}\left(\frac{\alpha_{ij}}{2}\left(1-\frac{f_{13}(r_{ij})}{r_{vdW}}\right)\right)\right\}
\end{eqnarray}
where $f_{13}(r_{ij}) = \left(r_{ij}^{p_{vdW}} + \gamma_{vdW}^{-p_{vdW}}\right)^{\frac{1}{p_{vdW}}}$ is a shielding term included to avoid excessive repulsive interactions between bonded atoms and atoms containing a valence angle (1--3 interactions), $D_{ij}$ is the depth of the potential well, $r_{vdW}$ is the van der Waal radius, $p_{vdW}$ and $\gamma_{vdW}$ are the van der Waals shielding parameters and $\mathcal{T}(r_{ij})$ is the Taper correction. The Coulomb interaction between atoms $i$ and $j$ is
\begin{equation}
E_{C,ij} = \mathcal{T}(r_{ij})C\frac{q_iq_j}{\left(r_{ij}^3 + \gamma_{ij}^{-3}\right)^{\frac{1}{3}}},
\end{equation}
where $q_i$ and $q_j$ are instantaneous charges of atoms $i$ and $j$, $C$ is the Coulomb constant and $\gamma_{ij}$ is a shielding parameter included to avoid excessive repulsions due to overlap of orbitals at short distances.

The atomic charges are calculated at every iteration (time step) during minimization (MD) run using the Electronegativity Equalization Method (EEM).\cite{Rappe1991} This bond-order formalism coupled with the redistribution of charges through EEM enables the ReaxFF model to describe ionic, metallic, and covalent systems on equal footing.\cite{Goddard2002,Han2005,Ojwang2008,Ojwang2009,VanDuin2001,VanDuin2003,Strachan2005,Buehler2006,Raymand2008,Zhang2004,Cheung2005}
With one unified (albeit complicated) formalism, ReaxFF has several advantages over other EFFs:\cite{vanDuin_book}

\begin{enumerate}
\item[(i)]{The bond-order formalism provides a continuous description of formation and dissociation of bonds during a molecular dynamics simulation.}
\item[(ii)]{Other interatomic potentials based on bond-order formalism like Tersoff\cite{Tersoff1988} and Brenner\cite{Benner1990} do not account for redistribution of charges. The EEM\cite{Rappe1991} employed in ReaxFF allows the atomic charges to vary continuously with changes in coordination and bond order.
\item[(iii)]{The evaluation of individual contributions of $\sigma$-, $\pi$- and double $\pi$- bonds to the bond order allows ReaxFF to identify the hybridization state and the coordination of an atom based on the instantaneous geometry around that atom.}
\item[(iv)]{The bond-order correction scheme enables ReaxFF to capture more accurately transition states during a reaction, provides a continuous transition between these intermediate states, and in turn describes reaction kinetics better than the other EFFs.}
}
\end{enumerate}

\subsection{ReaxFF development}\label{subsec:method}
The choice of specific partial energy contributions depends largely on the system
of interest. For example, for ionic solids the angle bending and torsion terms have been
set to zero;\cite{Goddard2002,Han2005,Ojwang2008,Ojwang2009} however, for
covalent crystals these contributions cannot be neglected as shown, {\em e.g.}, for the case of Si/SiO$_2$.\cite{VanDuin2003}
The partial energy contributions used in the development of ReaxFF for Si/SiO$_2$ system\cite{VanDuin2003} have
been found adequate in the present study of Li/Al/Si/O as well [Eq.~(\ref{eq:total_energy})].
We optimized all ReaxFF parameters for the Li/Al/Si/O system by fitting
against DFT-computed data. To ensure good transferability of the resulting Li/Al/Si/O parameters,
we have included in the training set DFT-calculated data for a wide variety of well-known
condensed phases and clusters, as listed below:
\begin{enumerate}
\item[(i)]{Equations of state ({\em i.e.} total energy versus volume) for pure Al (fcc, hcp, bcc, sc and diamond) and for corundum ($\alpha$-Al$_2$O$_3$), surface energy of the fcc Al (111), charge distribution and dissociation energies of a number of Al$-$O$-$H clusters; data from Ref.~\onlinecite{Zhang2004}.}
\item[(ii)]{Equations of state of Li (bcc, fcc, hcp, diamond, sc), LiH with sodium-chloride structure, dissociation energies and charge distributions in Li$_2$, LiH and LiH$_2$ clusters; data from Ref.~\onlinecite{Han2005}.}
\item[(iii)]{Equations of state of Si (sc, diamond, $\beta$-Sn), SiO$_2$ ($\alpha$-quartz, trydimite, coesite, $\alpha$-crystobalite, stishovite), dissociation energies of single and double bonds of Si$-$Si and Si$-$O in Si/O/H clusters, energies of various Si/O/H clusters
as a function of valence angles Si$-$O$-$Si, O$-$Si$-$O and Si$-$Si$-$Si and distortion energies of rings of Si/O/H clusters; data from Ref.~\onlinecite{VanDuin2003}.}
\item[(iv)]{Equations of state of Li-silicates: (a) Li$_2$SiO$_3$ (orthorhombic) (b) Stable Li$_2$Si$_2$O$_5$ (monoclinic) and (c) Metastable Li$_2$Si$_2$O$_5$ (orthorhombic); data from Ref.~\onlinecite{du}.}
\end{enumerate}

In addition to using DFT data sets from earlier works, we calculated the equations of state of the
following condensed phases within the framework of DFT using the computational details
listed in Sec.~\ref{subsec:comp_details}:
\begin{enumerate}
\item[(v)]{Li-oxides: $\alpha$-Li$_2$O (cubic)\cite{Lazicki2006} and Li$_2$O$_2$ (hexagonal).\cite{Cota2005}}
\item[(vi)]{Li-aluminates: Three polymorphs of LiAlO$_2$ namely, (a) $\alpha$ (rhombohedral)\cite{Marezio_alpha1965}, (b) $\beta$ (orthorhombic),\cite{Dronskowski1993} and (c) $\gamma$ (tetragonal).\cite{Wu2009}}
\item[(vii)]{Al-silicates: Three polymorphs of Al$_2$SiO$_5$ namely, (a) Andalusite (orthorhombic),\cite{Winter1979,Ralph1984} (b) Sillimanite (orthorhombic),\cite{Winter1979,Yang_PCM1997} and (c) Kyanite (triclinic).\cite{Winter1979,Yang_AmM1997}}
\end{enumerate}

In order to account for anisotropy, computational supercells of these phases were subjected to different
types of strain (depending on the crystal symmetry) when computing their energy as a function of cell volume.
The lattice of a crystal is described by three lattice vectors $\mathbf{a}_i$ ($i=1,2,3$) whose
magnitudes are the lattice parameters $a_i$. The cubic phases ($a_1$ = $a_2$ = $a_3$) were strained
triaxially, i.e., all the three lattice vectors ($a_i$) were all equally strained. The tetragonal and
hexagonal phases ($a_1$ = $a_2$ $\neq$ $a_3$) were deformed by two types of strains, namely (a)
biaxial: $a_1$ and $a_2$ were strained simultaneously by the same amount while keeping $a_3$ fixed
at its experimental value, and (b) uniaxial: $a_3$ was strained while keeping $a_1$ and $a_2$ fixed.
The lattice vectors of phases with orthorhombic symmetry or lower ($a_1$ $\neq$ $a_2$ $\neq$ $a_3$)
were strained individually keeping the other two unstrained, which leads to three distinct uniaxial
strains corresponding to three lattice vector directions $\mathbf{a}_i$. In all the cases, the
limits of strains range from $-$40\% (compressive) to $+$20\% (tensile).

For all the phases listed above [items (i)--(vii)], we computed the heats of formation $\Delta H_f$ as
functions of volume for the different types
of strains. The heat of formation of a general compound of unit formula (u.f.)
Li$_k$Al$_l$Si$_m$O$_n$ ($k,l,m,n$ integers $\geq 0$) at a
volume V for a particular type and value of strain can be evaluated from DFT total energy calculations as:
\begin{eqnarray}
\label{Eq:H_f}
\Delta H_f(V,\epsilon) &=& E_{\mathrm{Li}_k\mathrm{Al}_l\mathrm{Si}_m\mathrm{O}_n}(V,\epsilon) - kE_{\mathrm{Li}} \nonumber \\
                             & & -lE_{\mathrm{Al}}- mE_{\mathrm{Si}} -\frac{n}{2}E_{\mathrm{O}_2}
\end{eqnarray}
where $E_{\mathrm{Li}_k\mathrm{Al}_l\mathrm{Si}_m\mathrm{O}_n}$ is the total energy of a given volume V of the phase
Li$_k$Al$_l$Si$_m$O$_n$ subjected to a particular strain $\epsilon$. The energies of the constituent elements Li,
Al, Si and O in Eq.~(\ref{Eq:H_f}), {\em i.e.}, $E_{\mathrm{Li}}$, $E_{\mathrm{Al}}$, $E_{\mathrm{Si}}$, and $E_{\mathrm{O}_2}$, are those of
the most stable phases at equilibrium calculated by DFT.

The training set data were used to parameterize the ReaxFF using the successive one-parameter search technique described by van
Duin \emph{et al.}\cite{VanDuin1994} These parameters are tabulated in Appendix~\ref{appendix:param}, and are also made available as a data file.\cite{Text2010}

\begin{figure*}[htbp]
\begin{center}
\includegraphics[width=10.0cm]{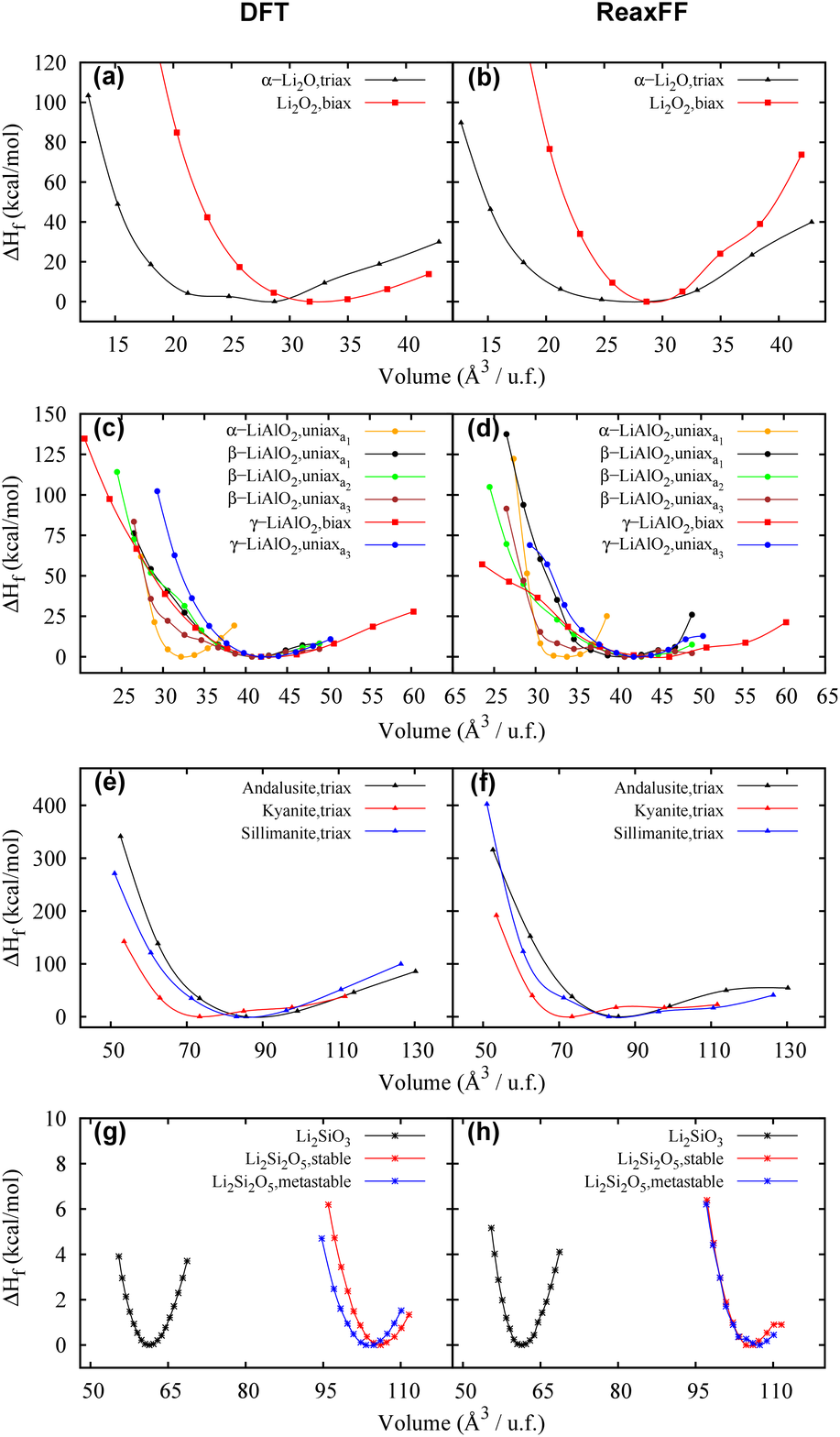}
\caption{(Color online) Equations of state of various phases of (a, b) Li oxides, (c, d) Li aluminates, (e, f) Al silicates, and (g, h) Li silicates as calculated
using DFT [panels (a), (c), (e), (g)] and ReaxFF [panels (b), (d), (f), (h)]. } \label{fig:comparison_hf}
\end{center}
\end{figure*}

\subsection{Details of the DFT calculations}
\label{subsec:comp_details}
The computational supercell for each phase in the training set described in Sec.~\ref{subsec:method} consisted of
one primitive unit cell. The total energy DFT calculations were performed within the framework of the generalized
gradient approximation (GGA), using the projector-augmented wave (PAW) formalism\cite{PAW} as implemented in the ab-initio
simulation package VASP.\cite{VASP1,VASP2} The atomic coordinates were relaxed using a conjugate gradient algorithm
until the force components on any atom were smaller than 0.01 eV/\AA. The exchange-correlation was described by the
Perdew-Wang functional,\cite{PW91} which is a typical choice for ceramics oxide systems ({\em e.g.}, Ref.~\onlinecite{du, Wu2009}).
The plane wave energy cutoff was set to 500 eV, which performs satisfactorily for similar ceramic systems.\cite{du}
The Brillouin (BZ) zone was sampled with a $\Gamma$-centered Monkhorst-Pack grid. For the oxides and aluminates of
lithium, we used $8\times8\times8$ $k$-point grids which amount to 1024 irreducible $k$-points for oxides and
256 $k$-points for aluminates. A $4\times4\times4$ $k$-point grid was found sufficient for the aluminum silicate
phases (32 irreducible $k$-points), and a $3\times3\times3$ $k$-point grid was selected for the eucryptite
phases (14 irreducible $k$-points). These grids were chosen on the basis of convergence tests
conducted for different BZ samplings for different phases.

\section{Results} \label{sec:results}

\subsection{Heats of formation} \label{subsec:ReaxFF_Development}
The set of parameters obtained by the technique described in Sec.~II were
validated by comparing the structures and the heats of formations for various phases
calculated by ReaxFF with those known from experiments or from DFT calculations. As
a preliminary test, the heats of formation as functions of volume for the various phases used in
the training set calculated by ReaxFF were compared in Fig.~\ref{fig:comparison_hf} with their DFT counterparts.
Figure~\ref{fig:comparison_hf} shows a generally good qualitative agreement between the ReaxFF and the DFT curves
in terms of equilibrium volumes and the relative phase stabilities at these volumes. Furthermore,
Table~\ref{table:comparison_hf} shows that the ReaxFF heat of formation results are also in good agreement with {\em experimental}
data on selected oxides, aluminates, and silicates at equilibrium.
However, in the deformation regimes lying outside equilibrium (particularly in tension) the energetic ordering of Li
oxides [Fig.~\ref{fig:comparison_hf}(a,b)], Li aluminates [Fig.~\ref{fig:comparison_hf}(c,d)], and Al
silicates [Fig.~\ref{fig:comparison_hf}(e,f)] at the ReaxFF level does not preserve so well the DFT ordering. This is most likely
due to the choice of deformation range (Sec.~\ref{subsec:method}), which contains more data
points in compression than in tension. The following subsections contain more
tests of the performance of ReaxFF concerning the structure, stability, and elastic properties of LAS ceramics.

\begin{table}
\caption{Heats of formation at equilibrium ($\Delta H_f^{\circ}$) of selected phases calculated using DFT and ReaxFF at 0 K.
For comparison, experimental values at 298 K are also provided wherever available.} \label{table:comparison_hf}
\begin{center}
\begin{tabular}{lcccccc}
\toprule
%&&&&&&\\
 Phase               &\hspace{0.65cm}  &  \multicolumn{5}{c}{$\Delta H_f^\circ$ (kcal mol$^{-1}$)} \\
\cline{3-7}
                     &\hspace{0.65cm}  &   DFT            &\hspace{0.65cm}  &  ReaxFF &\hspace{0.65cm}  &  Exp\\
\colrule
%                     &&         &&         &&                 \\
$\alpha$-Li$_2$O     &\hspace{0.65cm}  & -147.67 &\hspace{0.65cm}  & -145.82 &\hspace{0.65cm}  & -143.10$^a$         \\
%                     &&         &&         &&                 \\
$\gamma$-LiAlO$_2$   &\hspace{0.65cm}  & -292.19 &\hspace{0.65cm}  & -299.29 &\hspace{0.65cm}  & -284.37$^b$         \\
%             	     &&         &&         &&                 \\
Andalusite           &\hspace{0.65cm}  & -635.90 &\hspace{0.65cm}  & -635.35 &\hspace{0.65cm}  & -619.42$^c$         \\
%                     &&         &&         &&                 \\
Li$_2$SiO$_3$        &\hspace{0.65cm}  & -406.04 &\hspace{0.65cm}  & -387.31 &\hspace{0.65cm}  & -395.77$^d$         \\
%                     &&         &&         &&                 \\
$\alpha$-LiAlSiO$_4$ &\hspace{0.65cm}  & -529.25 &\hspace{0.65cm}  & -526.48 &\hspace{0.65cm}  & -512.53$^e$             \\
%                     &&         &&         &&                 \\
$\beta$-LiAlSiO$_4$  &\hspace{0.65cm}  & -528.01 &\hspace{0.65cm}  & -525.51 &\hspace{0.65cm}  & -506.18$^f$          \\
$\gamma$-LiAlSiO$_4$ &\hspace{0.65cm}  & -524.93 &\hspace{0.65cm}  & -517.66 &\hspace{0.65cm} &                 \\
\botrule
\multicolumn{7}{l}{\footnotesize{$^a$Ref.~\onlinecite{Chase1988}; $^b$Ref.~\onlinecite{Robie1995}; $^c$Ref.~\onlinecite{Stolen2004}; $^d$Ref.~\onlinecite{Nakagawa1981}; $^e$Ref.~\onlinecite{Fasshauer1998}; $^f$Ref.~\onlinecite{Heaney1999}}}\\
\end{tabular}
\end{center}
\end{table}

\subsection{Structural parameters}
Table~\ref{table:lattice_parameters} compares the lattice parameters for a number of selected phases calculated
using ReaxFF with those from DFT calculations and from experiments. These
lattice constants were calculated by optimizing the computational supercell of each phase
with respect to all independent lattice parameters that describe its crystal structure. Table~\ref{table:lattice_parameters} shows that
the lattice parameters calculated using ReaxFF are all within $\sim$5\% of the values reported in literature using
DFT or experiments. In order to establish that the structures of bulk phases are faithfully described by ReaxFF, we
have also checked the independent fractional coordinates of the atoms in the optimized supercells. For example, Tables~\ref{table:alpha_coordinates}
and \ref{table:beta_coordinates} show these fractional coordinates for $\alpha$- and $\beta$-eucryptite, respectively.
As shown in these tables, the agreement between the ReaxFF-predicted values for fractional coordinates and the experimental
ones is very good, which illustrates that ReaxFF predicts the structure of bulk phases accurately.

\begin{table*}
\caption{Comparison of calculated lattice parameters of selected phases using ReaxFF with those available
in literature determined from DFT calculations and from experiments. The numbers of unit formulae per
unit cell are provided in brackets.}\label{table:lattice_parameters}
\begin{center}
\begin{tabular}{lcccccccccccccc}
\toprule
%&&&&&&&&&&&&&&\\
Phase               && Structure  && Space Group                 && Formula  && Lattice    && DFT   && ReaxFF && Exp \\
                    &&                    &&                             &&                      && parameter (\AA)  &&       &&        &&     \\
\colrule
%                    &&                    &&                             &&                  &&            &&           &&        &&      \\
$\alpha$-Li$_2$O    && Cubic              && $Fm\overline{3}m$   && Li$_{2}$O (4)    && $a$    && 4.631$^a$ && 4.738  && 4.622$^b$ \\
%                    &&                    &&                             &&                  &&            &&           &&        &&       \\
\colrule
%                    &&                    &&                             &&                  &&            &&           &&        &&       \\
$\gamma$-LiAlO$_2$  && Tetragonal         && $P$4$_1$2$_1$2      && LiAlO$_2$ (4)    && $a$     && 5.223$^c$ && 5.359  && 5.169$^d$ \\
                    &&                    &&                             &&                  && $c$    && 6.309$^c$ && 6.234  && 6.268$^d$ \\
%                    &&                    &&                             &&                  &&            &&           &&        &&       \\
\colrule
%                    &&                    &&                             &&                  &&            &&           &&        &&       \\
Andalusite 	    && Orthorhombic       && $Pnnm$             && Al$_2$SiO$_5$ (4)&& $a$    && 7.753$^e$ && 7.632  && 7.798$^f$ \\
                    &&                    &&                             &&                  && $b$     && 7.844$^e$&& 7.916  && 7.903$^f$  \\
                    &&                    &&                             &&                  && $c$    && 5.477$^e$ && 5.727  && 5.557$^f$ \\
%                    &&                    &&                             &&                  &&            &&           &&        &&       \\
\colrule
%                    &&                    &&                             &&                  &&            &&           &&        &&       \\
Li$_2$SiO$_3$       && Orthorhombic       && $Cmc$2$_1$         && Li$_2$SiO$_3$ (4)&& $a$     && 9.487$^g$ && 9.335  && 9.392$^h$ \\
                    &&                    &&                             &&                  && $b$    && 5.450$^g$ && 5.431  && 5.397$^h$ \\
                    &&                    &&                             &&                  && $c$    && 4.713$^g$ && 4.861  && 4.660$^h$ \\
%                    &&                    &&                             &&                  &&            &&           &&        &&       \\
\colrule
%                    &&                    &&                             &&                  &&            &&           &&        &&       \\
$\alpha$-eucryptite && Trigonal           && $R$3               && LiAlSiO$_4$ (18) && $a$    &&  13.656    && 13.448    && 13.532$^i$\\
                    &&                    &&                    &&                  && $c$    &&   9.158    && 8.981    && 9.044$^i$ \\
%                    &&                    &&                             &&                  &&            &&           &&        &&       \\
\colrule
%                    &&                    &&                             &&                  &&            &&           &&        &&       \\
$\beta$-eucryptite  && Hexagonal          && $P$6$_4$22        && LiAlSiO$_4$ (12) && $a$    && 10.594$^j$&& 10.568 && 10.497$^k$\\
                    &&                    &&                             &&                  && $c$    && 11.388$^j$&& 11.763 && 11.200$^k$\\
%                    &&                    &&                             &&                  &&            &&           &&       &&         \\
\botrule
\multicolumn{15}{l}{\footnotesize{$^a$Ref.~\onlinecite{Lazicki2006}; $^b$Ref.~\onlinecite{Kunc2005}; $^c$Ref.~\onlinecite{Wu2009}; $^d$Ref.~\onlinecite{Marezio1965}; $^e$Ref.~\onlinecite{Winkler2001}; $^f$Ref.~\onlinecite{Winter1979}; $^g$Ref.~\onlinecite{Tang2010}; $^h$Ref.~\onlinecite{Hesse1977}; $^i$Ref.~\onlinecite{Daniels2001}; $^j$Ref.~\onlinecite{Narayanan2010}; $^k$Ref.~\onlinecite{Pillars1973}}}\\
\end{tabular}
\end{center}
\end{table*}

\begin{table}
\caption{Fractional coordinates of atoms in a unit cell of $\alpha$-eucryptite calculated using ReaxFF at 0 K. The experimental values from Ref.~\onlinecite{Daniels2001} at 298 K are provided for comparison.} \label{table:alpha_coordinates}
\begin{center}
\begin{tabular}{lcccccccccccc}
\toprule
%&&&&&&\\
        &\hspace{0.1cm} & \multicolumn{5}{c}{ReaxFF} &\hspace{0.1cm} & \multicolumn{5}{c}{Experiment}\\
\cline{3-7}  \cline{9-13}
%                    &&           &&           &&                 \\
Atom    &\hspace{0.1cm} & $x$       &\hspace{0.1cm} &   $y$     &\hspace{0.1cm} &    $z$    &\hspace{0.1cm} &    $x$    &\hspace{0.1cm}&   $y$  &\hspace{0.1cm} &     $z$     \\
%                    &&           &&           &&                 \\
\colrule
Li(1)	&\hspace{0.1cm} & -0.016         &\hspace{0.1cm} & -0.806         &\hspace{0.1cm} & -0.752       &\hspace{0.1cm} & -0.017         &\hspace{0.1cm} & -0.811         &\hspace{0.1cm} & -0.749      \\
%       &&           &&           &&           &&           &&           &&          \\
Li(2)	&\hspace{0.1cm} & 0.022       &\hspace{0.1cm} & 0.814         &\hspace{0.1cm} & 0.749         &\hspace{0.1cm} & 0.021       &\hspace{0.1cm} &0.812         &\hspace{0.1cm} & 0.754        \\
%       &&           &&           &&           &&           &&           &&          \\
Si(1)	&\hspace{0.1cm} & 0.531       &\hspace{0.1cm} & 0.876         &\hspace{0.1cm} & 0.753    &\hspace{0.1cm} & 0.530       &\hspace{0.1cm} & 0.880         &\hspace{0.1cm} & 0.750   \\
%       &&           &&           &&           &&           &&           &&          \\
Si(2)	&\hspace{0.1cm} & 0.876    &\hspace{0.1cm} & 0.348         &\hspace{0.1cm} & 0.918         &\hspace{0.1cm} & 0.876    &\hspace{0.1cm} & 0.344       &\hspace{0.1cm} & 0.916        \\
%       &&           &&           &&           &&           &&           &&          \\
Al(1)	&\hspace{0.1cm} & -0.533    &\hspace{0.1cm} & -0.883    &\hspace{0.1cm} & 0.754         &\hspace{0.1cm} & -0.530   &\hspace{0.1cm} & -0.882    &\hspace{0.1cm} & -0.749        \\
%       &&           &&           &&           &&           &&           &&          \\
Al(2)   &\hspace{0.1cm} & -0.878    &\hspace{0.1cm} & -0.342         &\hspace{0.1cm} & -0.914       &\hspace{0.1cm} & -0.875    &\hspace{0.1cm} & -0.345         &\hspace{0.1cm} & -0.916      \\
%       &&           &&           &&           &&           &&           &&          \\
O(1)	&\hspace{0.1cm} & -0.748    &\hspace{0.1cm} & -0.208    &\hspace{0.1cm} & -0.897       &\hspace{0.1cm} & -0.753    &\hspace{0.1cm} & -0.210   &\hspace{0.1cm} & -0.890      \\
%       &&           &&           &&           &&           &&           &&          \\
O(2)	&\hspace{0.1cm} & 0.764    &\hspace{0.1cm} & 0.211    &\hspace{0.1cm} & 0.903    &\hspace{0.1cm} & 0.766    &\hspace{0.1cm} & 0.216    &\hspace{0.1cm} & 0.898   \\
%       &&           &&           &&           &&           &&           &&          \\
O(3)    &\hspace{0.1cm} & -0.741    &\hspace{0.1cm} & -0.202    &\hspace{0.1cm} & -0.594    &\hspace{0.1cm} & -0.733    &\hspace{0.1cm} & -0.199    &\hspace{0.1cm} & -0.593   \\
%       &&           &&           &&           &&           &&           &&          \\
O(4)    &\hspace{0.1cm} & 0.734    &\hspace{0.1cm} & 0.198    &\hspace{0.1cm} & 0.571    &\hspace{0.1cm} & 0.733    &\hspace{0.1cm} & 0.199    &\hspace{0.1cm} & 0.576   \\
%       &&           &&           &&           &&           &&           &&          \\
O(5)	&\hspace{0.1cm} & -0.097    &\hspace{0.1cm} & -0.886    &\hspace{0.1cm} & -0.931    &\hspace{0.1cm} & -0.105    &\hspace{0.1cm} & -0.888    &\hspace{0.1cm} & -0.937   \\
%       &&           &&           &&           &&           &&           &&          \\
O(6)	&\hspace{0.1cm} & 0.090    &\hspace{0.1cm} & 0.879    &\hspace{0.1cm} & 0.947    &\hspace{0.1cm} & 0.096    &\hspace{0.1cm} & 0.881    &\hspace{0.1cm} & 0.946   \\
%       &&           &&           &&           &&           &&           &&          \\
O(7)	&\hspace{0.1cm} & -0.669    &\hspace{0.1cm} & -0.009    &\hspace{0.1cm} & -0.751   &\hspace{0.1cm} & -0.664    &\hspace{0.1cm} & -0.009    &\hspace{0.1cm} & -0.749   \\
%       &&           &&           &&           &&           &&           &&          \\
O(8)	&\hspace{0.1cm} & 0.656    &\hspace{0.1cm} & -0.004    &\hspace{0.1cm} & 0.753    &\hspace{0.1cm} & 0.655    &\hspace{0.1cm} & -0.004    &\hspace{0.1cm} & 0.750   \\
\botrule
\end{tabular}
\end{center}
\end{table}

\begin{table}
\caption{Fractional coordinates of atoms in a unit cell of $\beta$-eucryptite calculated using ReaxFF at 0 K. The experimental values from Ref.~\onlinecite{Xu1999} at 298 K are provided for comparison.} \label{table:beta_coordinates}
\begin{center}
\begin{tabular}{lcccccccccccc}
\toprule
%&&&&&&\\
        &\hspace{0.1cm} & \multicolumn{5}{c}{ReaxFF} &\hspace{0.1cm} & \multicolumn{5}{c}{Experiment}\\
\cline{3-7}  \cline{9-13}
%                    &&           &&           &&                 \\
Atom    &\hspace{0.1cm} & $x$       &\hspace{0.1cm} &   $y$     &\hspace{0.1cm} &    $z$    &\hspace{0.1cm} &    $x$    &\hspace{0.1cm}&   $y$  &\hspace{0.1cm} &     $z$     \\
%                    &&           &&           &&                 \\
\colrule
Li(1)	&\hspace{0.1cm} & 0.000         &\hspace{0.1cm} & 0.000         &\hspace{0.1cm} & 0.500       &\hspace{0.1cm} & 0.000         &\hspace{0.1cm} & 0.000         &\hspace{0.1cm} & 0.500      \\
%       &&           &&           &&           &&           &&           &&          \\
Li(2)	&\hspace{0.1cm} & 0.500       &\hspace{0.1cm} & 0.000         &\hspace{0.1cm} & 0.000         &\hspace{0.1cm} & 0.500       &\hspace{0.1cm} & 0.000         &\hspace{0.1cm} & 0.000        \\
%       &&           &&           &&           &&           &&           &&          \\
Li(3)	&\hspace{0.1cm} & 0.500       &\hspace{0.1cm} & 0.000         &\hspace{0.1cm} & 0.327    &\hspace{0.1cm} & 0.500       &\hspace{0.1cm} & 0.000         &\hspace{0.1cm} & 0.328   \\
%       &&           &&           &&           &&           &&           &&          \\
Si(1)	&\hspace{0.1cm} & 0.248    &\hspace{0.1cm} & 0.000         &\hspace{0.1cm} & 0.000         &\hspace{0.1cm} & 0.248    &\hspace{0.1cm} & 0.000         &\hspace{0.1cm} & 0.000        \\
%       &&           &&           &&           &&           &&           &&          \\
Si(2)	&\hspace{0.1cm} & 0.251    &\hspace{0.1cm} & 0.502    &\hspace{0.1cm} & 0.000         &\hspace{0.1cm} & 0.247    &\hspace{0.1cm} & 0.494    &\hspace{0.1cm} & 0.000        \\
%       &&           &&           &&           &&           &&           &&          \\
Al(1)   &\hspace{0.1cm} & 0.258    &\hspace{0.1cm} & 0.000         &\hspace{0.1cm} & 0.500       &\hspace{0.1cm} & 0.250    &\hspace{0.1cm} & 0.000         &\hspace{0.1cm} & 0.500      \\
%       &&           &&           &&           &&           &&           &&          \\
Al(2)	&\hspace{0.1cm} & 0.250    &\hspace{0.1cm} & 0.499    &\hspace{0.1cm} & 0.500       &\hspace{0.1cm} & 0.251    &\hspace{0.1cm} & 0.501    &\hspace{0.1cm} & 0.500      \\
%       &&           &&           &&           &&           &&           &&          \\
O(1)	&\hspace{0.1cm} & 0.115    &\hspace{0.1cm} & 0.201    &\hspace{0.1cm} & 0.248    &\hspace{0.1cm} & 0.112    &\hspace{0.1cm} & 0.199    &\hspace{0.1cm} & 0.242   \\
%       &&           &&           &&           &&           &&           &&          \\
O(2)    &\hspace{0.1cm} & 0.101    &\hspace{0.1cm} & 0.696    &\hspace{0.1cm} & 0.263    &\hspace{0.1cm} & 0.097    &\hspace{0.1cm} & 0.699    &\hspace{0.1cm} & 0.259   \\
%       &&           &&           &&           &&           &&           &&          \\
O(3)    &\hspace{0.1cm} & 0.604    &\hspace{0.1cm} & 0.704    &\hspace{0.1cm} & 0.262    &\hspace{0.1cm} & 0.597    &\hspace{0.1cm} & 0.705    &\hspace{0.1cm} & 0.264   \\
%       &&           &&           &&           &&           &&           &&          \\
O(4)	&\hspace{0.1cm} & 0.605    &\hspace{0.1cm} & 0.202    &\hspace{0.1cm} & 0.258    &\hspace{0.1cm} & 0.608    &\hspace{0.1cm} & 0.201    &\hspace{0.1cm} & 0.249   \\
\botrule
\end{tabular}
\end{center}
\end{table}

%%%%%%%%%%%

\subsection{Stability of eucryptite phases}
\begin{figure}[htbp]
\begin{center}
\includegraphics[width=6.0cm]{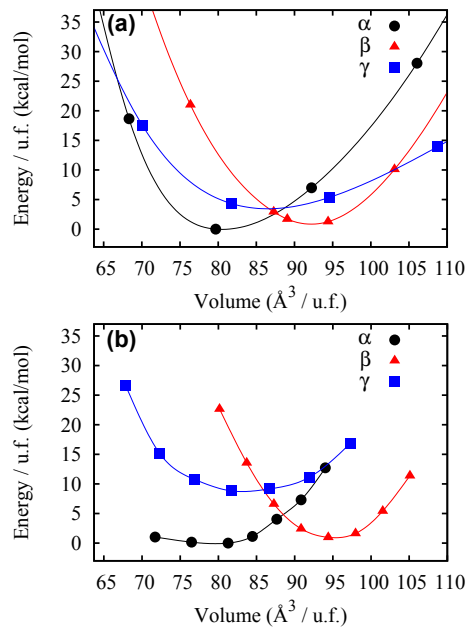}
\caption{(Color online) Equations of state of various phases of eucryptite
calculated using (a) DFT and (b) ReaxFF. Both the techniques predict
the same order of stability of the three polymorphs with $\alpha$ being the
most stable phase in each case.} \label{fig:order_stability_eucryptite}
\end{center}
\end{figure}

There are three well-known crystalline polymorphs of eucryptite, $\alpha$, $\beta$, $\gamma$; of these, $\alpha$ is the most stable phase under ambient conditions but is kinetically hindered.\cite{Beall} Fig.~\ref{fig:order_stability_eucryptite} shows the equations of state (Energy vs Volume curves) for these polymorphs calculated using ReaxFF and DFT at 0 K. The minimum of a calculated energy vs volume curve for a given phase represents its equilibrium state. For convenience, all the energies reported in Fig.~\ref{fig:order_stability_eucryptite} are given relative to the energy of the most stable phase at its equilibrium volume for DFT and ReaxFF. Fig.~\ref{fig:order_stability_eucryptite} shows that ReaxFF predicts the same order of stability for the three polymorphs of eucryptite as do our DFT calculations. This order is consistent with experimental observations.~\cite{Pillars1973,Daniels2001,Xu1999,Zhang2002}

$\beta$-eucryptite, the most technologically relevant of the three polymorphs, has a open structure which collapses at sufficiently high applied pressures.\cite{Bach,Palmer,Jochum2009} Recently, it was observed that $\beta$-eucryptite begins to amorphize at pressures above $\sim$5 GPa.\cite{Zhang2005} To test the ability of ReaxFF to capture phase transitions, we have studied the evolution of $\beta$-eucryptite under a rigid spherical indenter using MD simulations.\cite{LAMMPS} An orthorhombic simulation box of dimensions 41.99\AA\ $\times$ 72.73\AA\  $\times$ 56\AA\  containing 13440 atoms was used to simulate the crystal, which was indented down the $z$ axis (i.e., the [00$\overline{1}$] crystal direction). Periodic boundary conditions were applied in the directions perpendicular to the indentation force. The atoms that have $z$ coordinates within 12\AA\ of the lowest $z$ value (of all atoms) were kept fixed during MD runs in order to simulate the underlying bulk. The initial structure was relaxed at 0 K, and then thermalized at 300 K for 30 ps; the time-step used in the MD runs was 1 fs. After thermalization, the top face was indented at a rate of 0.065\AA/ps by a rigid spherical indenter of radius $R=14$\AA\ which applies a radial force $F_i$ on atom $i$ given by:
\begin{equation}
\label{Eq:force_indent}
F_i = \begin{cases} -k(r_i - R)^2 & \mbox{if } r_i \leq R \\
                     0            & \mbox{if } r_i > R
      \end{cases}
\end{equation}
where $k$ is a force constant ($k$ = 76.32 kcal/mol \AA$^3$), and $r_i$ is the distance between the center of the atom $i$ and that of the indenter.

During indentation simulations, we have not found any new phase at indent pressures smaller than 7 GPa even though the $\epsilon$ phase has been reported\cite{Zhang2002} to occur at $\sim$0.8 GPa. One reason for which we do not observe the $\epsilon$ phase in these
simulations is that the pressure is not applied hydrostatically (as it was in experiments\cite{Zhang2002}), and such anisotropic application of external pressure may lead to different phase transitions,\cite{m2} {\em i.e.,} different onset pressure or different phases. We have observed the formation of a denser but disordered phase in the vicinity of the indenter. Zhang {\em et al.}\cite{Zhang2005} carried out in-situ X-ray experiments on polycrystalline samples of $\beta$ eucryptite and reported that amorphization begins at a pressure $\sim$5 GPa and completes at a pressure of 17 GPa.  Our indentation MD simulations predict a higher onset pressure for amorphization, $\sim$7 GPa. However, it should be noted that the simulations were carried out on single crystal $\beta$ eucryptite; this eliminates the defects, porosity, or grain boundaries from our starting phase which would have acted as nucleation sites for the formation of the amorphous phase. Consequently, in the case of our simulations, there is an increased barrier towards amorphization, which is reflected in the increased onset pressure. At the ReaxFF onset pressure of 7 GPa, only regions near the indent amorphize, while those far away from it remain crystalline. As the indent pressure is increased, the amorphized region grows and there is a range of pressures over which amorphization proceeds: this finding is similar to what of Zhang {\em et al.} have found for polycrystalline samples.\cite{Zhang2005}

\begin{figure}[htbp]
\begin{center}
\includegraphics[width=5.5cm]{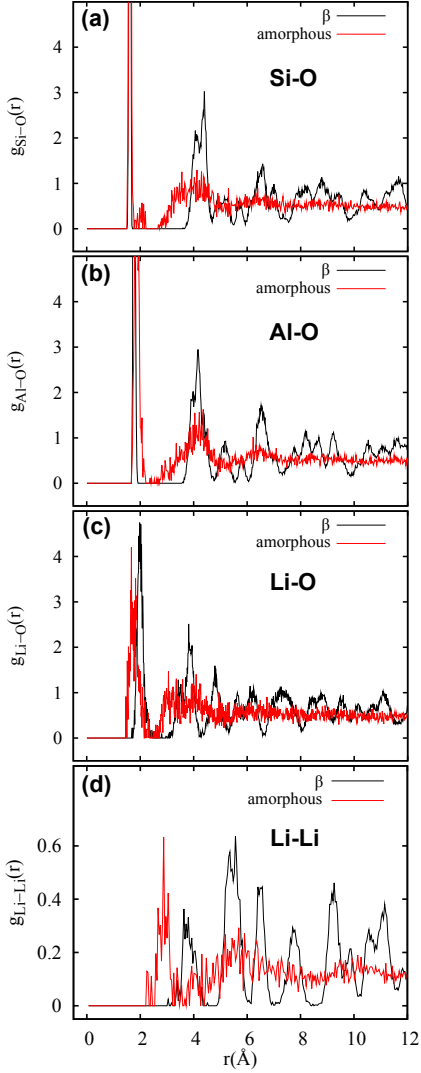}
\caption{(Color online) Pair distribution functions ($g_{\mathrm{A-B}}(r)$) for (a) Si--O, (b) Al--O, (c) Li--O and (d) Li--Li pairs in $\beta$-eucryptite (black lines)  and in the phase obtained under a spherical indent (red lines) at an applied contact pressure $\sim$ 10 GPa. The broadening of the peaks corresponding to higher order neighbors and lowering of the nearest-neighbor distances in Li--O and Li--Li pairs at high pressures indicates that the new phase formed under the indent is amorphous.} \label{fig:amorphization}
\end{center}
\end{figure}

We have analyzed the amorphous phase by studying radial distribution functions (RDF) for pairs of different types of atoms in the disordered region.
Specifically, the RDFs $g_{\mathrm{A-B}}(r)$ were evaluated for Si--O, Al--O, Li--O and Li--Li pairs after indentation proceeded to different depths $h$. Fig.~\ref{fig:amorphization} shows RDFs evaluated prior to the indentation, compared to those calculated after indentation to $h$ = 12\AA; at this depth, we evaluated the contact pressure at $\sim$10 GPa. The RDFs for all the type pairs considered show, prior to indentation, well-defined peaks at characteristic distances of $\beta$-eucryptite. Under pressure, the first peak ($r=1.6$\AA) of $g_{\mathrm{Si-O}}(r)$ [see Fig.~\ref{fig:amorphization}(a)] broadens somewhat and decreases in intensity compared to that of the crystalline $\beta$ phase. The peak corresponding to the second-nearest neighbor ($r=4.1$\AA) is very broad, while peaks at higher distances are not defined. A similar behavior of the RDF was observed in the amorphous phase obtained from high pressure MD simulations of $\beta$-crystobalite (SiO$_2$).\cite{Zhang1993} Fig.~\ref{fig:amorphization}(b) shows the RDF for Al--O, which also exhibits the tell-tale signs of a disordered phase under pressure; the broadening of the first peak, along with the disappearance of the higher-order peaks, has also been observed during amorphization of SiC\cite{Rino2004} and $\alpha$-quartz.\cite{Watson1995} Significant changes in the RDFs occur at contact pressures $\ge$7 GPa and indicate amorphization, which is also apparent from direct visualization of the structure. The pressure necessary for the onset of amorphization is consistent with empirical observations.\cite{Zhang2005}

Interestingly, an additional feature is exhibited by the RDFs of Li--O and Li--Li pairs [Fig.~\ref{fig:amorphization}(c,d)]. The first peaks for the Li--Li and Li--O pairs are shifted significantly to lower distances in the high pressure phase as compared to the initial crystal [Fig.~\ref{fig:amorphization}(c,d)]. For the Li--O pairs, the first peak shifts from 2\AA \  to 1.67\AA\ under pressure [Fig.~\ref{fig:amorphization}(c)], which is close to the typical Li--O bond length of 1.606\AA.\cite{Bellert2001} Fig.~\ref{fig:amorphization}(d) shows that the smallest most probable Li--Li spacing at high pressure is $\sim$ 2.88\AA\, which is closer to the experimental value of the bond length (3.04\AA) in Li-metal\cite{Weast1987} than it is to the lowest Li-Li distance (3.8\AA) in $\beta$-eucryptite. This suggests the existence of Li--Li bonds in the high pressure phase, which were not present in the crystalline phase; by checking the atomic structure details of the amorphized
phase, we have confirmed the presence of direct Li--Li bonds and have also found bonds in which two Li atoms are "bridged" by on O atom in a triangular configuration. These newly formed, compressed Li--Li bonds and the shortened Li-O bonds formed under pressure suggest densification in the amorphous phase.

%%%%%%%%%%%

\subsection{Elastic properties of eucryptite phases}

To assess the performance of ReaxFF in predicting elastic properties, we computed
the elements of elastic stiffness tensor $C_{ij}$ for two polymorphs
of eucryptite, $\alpha$ and $\beta$, by employing the technique outlined in
Appendix~\ref{appendix:elastic_calc}. We have found that the stiffness tensors for
both eucryptite phases are positive definite, which means that at the ReaxFF level
the Born stability criterion\cite{Born1956} is met. The seven independent elastic constants of
$\alpha$-eucryptite (rhombohedral structure) were calculated using ReaxFF at 0 K and
are listed in Table~\ref{table:a-eucryptite}; to the best of our knowledge, so far there are
no reports of elastic constants in the literature for this phase.

\begin{table}
\caption{Predicted stiffness constants ${C}_{ij}$ (in GPa)  of $\alpha$-eucryptite using ReaxFF at 0 K.}\label{table:a-eucryptite}
\begin{center}
\squeezetable
\begin{tabular}{lcccccccccccccc}
\toprule
%&&&&&&&&&&&&&&\\
          &&  $C_{11}$ &&  $C_{12}$ &&  $C_{13}$ &&  $C_{24}$ && $C_{15}$ && $C_{33}$ &&  $C_{44}$\\
\colrule
%          &&        &&        &&       &&        &&   &&  &&       \\
ReaxFF    && 131.86 && 67.92  && 25.18 && 1.96 && 1.28 && 175.24 && 37.29    \\
%          &&        &&        &&       &&        &&    &&  &&       \\
\botrule
\end{tabular}
\end{center}
\end{table}

We have also calculated the five independent elastic constants of $\beta$-eucryptite (hexagonal structure)
at 0 K predicted by ReaxFF and listed them in Table~\ref{table:b-eucryptite}. Hauss{\"u}hl \emph{et al.}\cite{Haussuhl1984}
have measured these elastic constants using an ultrasonic technique at ambient
temperature, 293 K. To compare the elastic constants predicted by ReaxFF in our study and
those predicted by DFT (also at 0 K)\cite{Narayanan2010} with the experimental values, we have extrapolated
the measured values of $C_{ij}$ to 0 K using thermoelastic constants $T_{ij} = d\mbox{log} C_{ij}/dT$.\cite{Haussuhl1984}
As shown in Table~\ref{table:b-eucryptite}, the ReaxFF elastic constants are in good agreement with experiment;
with the exception of $C_{12}$, all the calculated constants are within $\sim$ 30\% of experimental values
extrapolated to 0 K. These ReaxFF values are also consistent with those predicted by DFT and reported in
an earlier work.\cite{Narayanan2010}

\begin{table}
\caption{Comparison of the calculated stiffness constants ${C}_{ij}$ (in GPa) of $\beta$-eucryptite at 0 K with the experimental data from Ref.~\onlinecite{Haussuhl1984} extrapolated to 0 K using the thermoelastic constants $T_{ij} = d\log C_{ij}/dT$. The uncertainty in any of the experimental values (Exp) is smaller than 2.5 GPa.} \label{table:b-eucryptite}
\begin{center}
\begin{tabular}{lcccccccccc}
\toprule
%&&&&&&&&&&\\
          &&  $C_{11}$ &&  $C_{12}$ &&  $C_{13}$ &&  $C_{33}$ &&  $C_{44}$\\
\colrule
%     &&        &&        &&       &&        &&          \\
DFT$^a$  && 165.64 && 70.98  && 78.59 && 132.83 && 58.68    \\
%     &&        &&        &&       &&        &&          \\
ReaxFF   && 178.92 && 102.77  && 118.28 && 181.26 && 47.37    \\
%     &&        &&        &&       &&        &&          \\
Exp      && 176.3 && 68.5  && 89.8 && 139.9 && 61.2    \\
%     &&        &&        &&       &&        &&          \\
$T_{ij}$ $(10^{-3}/K)$ && -0.14 && 0.13 && -0.27 && -0.42 && -0.24   \\
%     &&        &&        &&       &&        &&          \\
\botrule
$^a$\footnotesize{Ref.~\onlinecite{Narayanan2010}, 0 K}\\
\end{tabular}
\end{center}
\end{table}

The $\alpha$ and $\beta$ polymorphs of eucryptite are known to possess highly anisotropic physical properties.
The overall elastic anisotropy of hexagonal and rhombohedral crystals is usually assessed through three ratios,
$C_{11}/C_{33}$, $C_{12}/C_{13}$ and $2C_{44}/(C_{11}-C_{12})$, whose deviations from unity serve as
measures of the anisotropy in the crystals being studied.
Table~\ref{table:anisotropy} lists the anisotropy ratios for rhombohedral $\alpha$ and hexagonal $\beta$-eucryptite using the $C_{ij}$
predicted by ReaxFF.

%
%Although these ratios are in reasonable agreement with those derived using the DFT or experimental elastic constants (Table~\ref{table:anisotropy}), %we note that the $C_{11}/C_{33}$ of $\beta$ eucryptite given by ReaxFF is subunitary and smaller than its experimental and DFT counterparts. The %reason why this ratio is too small is that the ReaxFF $C_{33}$ is larger than the experimental value by about 40 GPa (or 29\%) but $C_{11}$ is very %close the experiments, which leads to a subunitary $C_{11}/C_{33}$. The large ReaxFF $C_{33}$ leads directly to a bulk modulus that is larger by %$\sim$30\% higher than the experimental value for beta-eucryptite. On the other hand, for $\alpha$ eucryptite the calculated ReaxFF, %crystalline-averaged (75.06 GPa) is much closer to the experimental value (74 GPa):\cite{Fasshauer1998} this suggests that the ReaxFF $C_{33}$ for %$\alpha$ eucryptite is less likely to be deviate significantly from the still-undetermined experimental value.
%

\begin{table}
\caption{Anisotropic factor ratios for $\alpha$ and $\beta$ eucryptite (LiAlSiO$_4$) evaluated using single crystal elastic constants ($C_{ij}$) predicted by ReaxFF in the present study. For $\beta$-eucryptite, these ratios are also calculated using $C_{ij}$ known by DFT and experiments for comparison.} \label{table:anisotropy}
\begin{center}
\begin{tabular}{ccccccccc}
\toprule
\noalign{\smallskip}
%                     &\hspace{0.35cm}  &  $\alpha$-LiAlSiO$_4$ &\hspace{0.35cm}  &\multicolumn{5}{c}{$\beta$-LiAlSiO$_4$} \\
%\cline{3-3} \cline{5-9}
Phase                 &\hspace{0.25cm}  & Technique &\hspace{0.25cm}  & $\displaystyle{\frac{C_{11}}{C_{33}}}$  &\hspace{0.25cm}  & $\displaystyle{\frac{C_{12}}{C_{13}}}$ &\hspace{0.25cm} &$\displaystyle{\frac{2C_{44}}{C_{11}-C_{12}}}$\\
\noalign{\smallskip}
%&&&&&&\\
\colrule
%                      &&         &&         &&                    \\
%                      &\hspace{0.25cm}  &        &                 &         &                 &                \\
$\alpha$-LiAlSiO$_4$  &\hspace{0.25cm}  & ReaxFF                &\hspace{0.25cm}  & 0.7525 &\hspace{0.25cm}  &  2.6974 &\hspace{0.25cm}  & 1.1664         \\
\noalign{\smallskip}
%                     &&         &&         &&      &&            \\
\colrule
%                      &&         &&         &&                     \\
%$\beta$-LiAlSiO$_4$   &                 &        &                 &         &                 &                \\
%&&&&&&&& \\
\noalign{\smallskip}
                      &\hspace{0.25cm}  & DFT$^a$               &\hspace{0.25cm}  & 1.2470 &\hspace{0.25cm}  &  0.9032 &\hspace{0.25cm}  & 1.2398         \\
%             	      &&         &&         &&      &&           \\
$\beta$-LiAlSiO$_4$   &\hspace{0.25cm}  & ReaxFF                &\hspace{0.25cm}  & 0.9871 &\hspace{0.25cm}  &  0.8688 &\hspace{0.25cm}  & 1.2441         \\
%                     &&         &&         &&      &&           \\
                      &\hspace{0.25cm}  & Exp$^b$               &\hspace{0.25cm}  & 1.2602 &\hspace{0.25cm}  &  0.7628 &\hspace{0.25cm}  & 1.1354         \\
\botrule
\multicolumn{9}{l}{\footnotesize{$^a$From $C_{ij}$ in Ref.~\onlinecite{Narayanan2010}}}\\
\multicolumn{9}{l}{\footnotesize{$^b$From $C_{ij}$ in Ref.~\onlinecite{Haussuhl1984}, extrapolated to 0 K.}}\\
%\multicolumn{7}{l}{\footnotesize{$^a$Ref.~\onlinecite{Chase1988}; $^b$Ref.~\onlinecite{Robie1995}; $^c$Ref.~\onlinecite{Stolen2004}; $^d$Ref.~\onlinecite{Nakagawa1981}; $^e$Ref.~\onlinecite{Fasshauer1998}; $^f$Ref.~\onlinecite{Heaney1999}}}\\
\end{tabular}
\end{center}
\end{table}

\begin{figure}[htbp]
\begin{center}
\includegraphics[width=5.5cm]{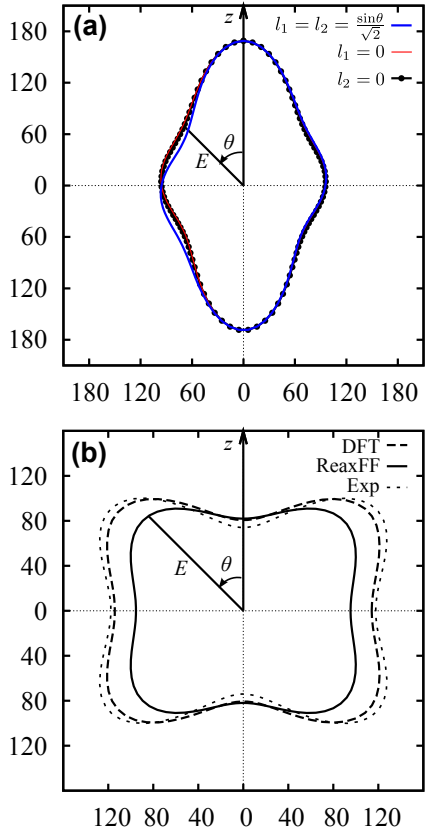}
\caption{(Color online) Polar plots illustrating the directional dependence of Young's Modulus $E$
for (a) $\alpha$-eucryptite using the elastic constants predicted by ReaxFF in
three different crystallographic planes containing the $z$-axis and (b) $\beta$-eucryptite using the
elastic constants predicted by ReaxFF and those known by DFT and experiments.}\label{fig:youngsmodulus_polar}
\end{center}
\end{figure}

The anisotropy of eucryptite  polymorphs manifests, expectedly, in the Young's modulus $E$ as well as in other elastic properties.
The direction-dependence of Young's modulus can be derived from the elastic constants, and we show it here as a way to directly visualize the anisotropic character of the Young's modulus (Fig.~\ref{fig:youngsmodulus_polar}).
The Young's modulus for a rhombohedral crystal in the $R3$ space group along a crystallographic direction of direction
cosines $l_1$, $l_2$, $l_3$ can be expressed in terms of elastic compliance constants as\cite{Nye}
\begin{eqnarray}
\label{eq:E_rhombo}
\frac{1}{E} & = & (1 - l_3^2)^2S_{11} + l_3^4S_{33} + l_3^2(1-l_3^2)(2S_{13}+S_{44})\nonumber  \\
            &   & + 2l_2l_3(3l_1^2-l_2^2)S_{14} + 2l_1l_3(3l_2^2-l_1^2)S_{25},
\end{eqnarray}
where $S_{ij}$ are the elements of elastic compliance matrix, $\mathbf{S}$ given by the inverse of the
elastic stiffness matrix, $\mathbf{C}$ \emph{i.e.}, $\mathbf{S} = \mathbf{C}^{-1}$. For hexagonal crystals, the directional
dependence of $E$ is given by\cite{Nye}
\begin{equation}
\label{eq:E_hexagonal}
%\frac{1}{E} & = & (1 - l_3^2)^2S_{11} + l_3^4S_{33} \nonumber \\
%            &   &   + l_3^2(1-l_3^2)(2S_{13}+S_{44})
\frac{1}{E}  = (1 - l_3^2)^2S_{11} + l_3^4S_{33} + l_3^2(1-l_3^2)(2S_{13}+S_{44})
\end{equation}

Figure~\ref{fig:youngsmodulus_polar}(a) shows the variation of Young's modulus of $\alpha$-eucryptite with the
angle $\theta$ between a given crystallographic direction and the $z$-axis, for three
different planes containing the $z$-axis; these planes are $yz$ ($l_1=0$), $xz$ ($l_2=0$) and the plane
containing the first bisector of the $xy$ plane, $l_1=l_2={\sin \theta}/\sqrt{2}$. The three polar
plots in Fig.~\ref{fig:youngsmodulus_polar}(a) were generated using Eq.~(\ref{eq:E_rhombo}) and the ReaxFF elastic constants
in Table~\ref{table:a-eucryptite}.

The Young's modulus of $\beta$-eucryptite depends only on the angle $\theta$ between a given direction
and the $z$-axis (crystallographic $c$-axis), owing to the symmetry of a hexagonal crystal. Figure~\ref{fig:youngsmodulus_polar} shows the
dependence of Young's modulus of $\beta$-eucryptite on $\theta$ calculated using the ReaxFF elastic constants
and those known from DFT and experiments (Table~\ref{table:b-eucryptite}). The variation with $\theta$ of the Young's modulus of $\beta$-eucryptite
calculated using elastic constants predicted by ReaxFF follows the same trends as
the $E$ calculated using the elastic constants known by DFT or experiments. We now focus on elastic properties corresponding to polycrystalline eucryptite phases.

\begin{table}
\caption{Average values of bulk moduli ($B$, in GPa) and shear moduli ($G$, in GPa) using the
Voigt, Ruess, and Hill's approximations for polycrystalline eucryptite phases
derived from their single-crystal elastic constants $C_{ij}$. The Young's moduli ($E_{poly}$, in GPa) and Poisson ratios
($\nu_{poly}$) are evaluated using Eqs.~(\ref{eq:Youngs}).} \label{table:polycrystalline}
\begin{center}
\begin{tabular}{lcccccccc}
\toprule
%&&&&&&\\
                     &\hspace{0.55cm}  &  $\alpha$-LiAlSiO$_4$ &\hspace{0.55cm}  &\multicolumn{5}{c}{$\beta$-LiAlSiO$_4$} \\
\cline{3-3} \cline{5-9}
                     &\hspace{0.55cm}  &   ReaxFF              &\hspace{0.55cm}  &  DFT$^a$ &\hspace{0.55cm}  &  ReaxFF  &\hspace{0.55cm} & Exp$^b$\\
\colrule
%                     &&         &&         &&                 \\
$B_V$                &\hspace{0.55cm}  &  75.06 &\hspace{0.55cm}  &  102.27 &\hspace{0.55cm}  & 135.31 &\hspace{0.55cm}  & 109.85         \\
%                     &&         &&         &&                 \\
$B_R$                &\hspace{0.55cm}  &  75.06 &\hspace{0.55cm}  &  101.55 &\hspace{0.55cm}  & 134.88 &\hspace{0.55cm}  & 109.55       \\
%             	     &&         &&         &&                 \\
$B_H$                &\hspace{0.55cm}  &  75.06 &\hspace{0.55cm}  &  101.91 &\hspace{0.55cm}  & 135.09 &\hspace{0.55cm}  & 109.70       \\
%                     &&         &&         &&                 \\
$G_V$                &\hspace{0.55cm}  &  42.69 &\hspace{0.55cm}  &   48.67 &\hspace{0.55cm}  &  39.88 &\hspace{0.55cm}  &  51.55     \\
%                     &&         &&         &&                 \\
$G_R$                &\hspace{0.55cm}  &  38.44 &\hspace{0.55cm}  &   46.08 &\hspace{0.55cm}  &  38.48 &\hspace{0.55cm}  &  47.10          \\
%                     &&         &&         &&                 \\
$G_H$                &\hspace{0.55cm}  &  40.56 &\hspace{0.55cm}  &   47.37  &\hspace{0.55cm} &  39.18 &\hspace{0.55cm}  &  49.32       \\
$E_{poly}$                  &\hspace{0.55cm}  & 103.12 &\hspace{0.55cm}  &  123.05 &\hspace{0.55cm}  & 107.18 &\hspace{0.55cm}  & 128.69              \\
$\nu_{poly}$                &\hspace{0.55cm}  &   0.27 &\hspace{0.55cm}  &    0.30 &\hspace{0.55cm}  &   0.37 &\hspace{0.55cm}  &   0.31          \\
\botrule
\multicolumn{9}{l}{\footnotesize{$^a$From $C_{ij}$ in Ref.~\onlinecite{Narayanan2010}}}\\
\multicolumn{9}{l}{\footnotesize{$^b$From $C_{ij}$ in Ref.~\onlinecite{Haussuhl1984}, extrapolated to 0 K.}}\\
%\multicolumn{7}{l}{\footnotesize{$^a$Ref.~\onlinecite{Chase1988}; $^b$Ref.~\onlinecite{Robie1995}; $^c$Ref.~\onlinecite{Stolen2004}; $^d$Ref.~\onlinecite{Nakagawa1981}; $^e$Ref.~\onlinecite{Fasshauer1998}; $^f$Ref.~\onlinecite{Heaney1999}}}\\
\end{tabular}
\end{center}
\end{table}

The theoretical average bulk ($B$) and shear ($G$) elastic moduli of polycrystalline $\alpha$ and $\beta$-eucryptite
can be derived from their single-crystal elastic constants. There are two well-known approximations
typically used to evaluate the polycrystalline elastic moduli, namely the
Voigt\cite{Voigt1928} and Reuss\cite{Reuss1929} methods, which provide upper bounds (identified by the subscript $V$) and lower bounds (subscript $R$), respectively, for the bulk and shear moduli. For rhombohedral and hexagonal crystal systems, the polycrystalline bulk and shear moduli can be expressed in terms of the single-crystal elastic constants as
\begin{eqnarray}\label{eq:VR}
  B_V         &=& \frac{1}{9}\left ( 2C_{11} + C_{33} + 2C_{12} + 4C_{13} \right )  \\
\frac{1}{B_R} &=& 2S_{11} + 2S_{12} + 4S_{13} + S_{33} \\
  G_V         &=& \frac{1}{30}\left ( 7C_{11} + 2C_{33} - 5C_{12} - 4C_{13} + 12C_{44} \right ) \\
\frac{1}{G_R} &=& \frac{1}{15}\left ( 14S_{11} - 10S_{12} - 8S_{13} + 4S_{33} + 6S_{44} \right )
\end{eqnarray}
The Hill values ($B_H$, $G_H$) of the bulk and shear moduli are the arithmetic averages
of the corresponding Voigt and Ruess bounds, and are considered the best estimates of these polycrystalline
moduli.\cite{Hill1952} The polycrystalline Young's modulus $E_{poly}$ and Poisson's ratio ($\nu_{poly}$) can be obtained through
the relations applicable to isotropic materials,\cite{Simmons1971}
\begin{equation}\label{eq:Youngs}
E_{poly} = \frac{9B_HG_H}{3B_H+G_H}; \quad \nu_{poly} = \frac{3B_H - 2G_H}{2(3B_H + G_H)}.
\end{equation}
Table~\ref{table:polycrystalline} lists the average elastic moduli of polycrystalline $\alpha$ and $\beta$ eucryptite
derived from the single-crystal constants $C_{ij}$ (Tables~\ref{table:a-eucryptite} and \ref{table:b-eucryptite})
through the relationships in Eqs.~(\ref{eq:VR})--(\ref{eq:Youngs}). For $\alpha$-eucryptite, the
polycrystalline bulk modulus evaluated using the single crystal elastic constants predicted
by ReaxFF (75.06 GPa) is in excellent agreement with an earlier measurement (74 GPa) of
Fasshauer \emph{et al.}\cite{Fasshauer1998} Furthermore, the polycrystalline elastic constants
of $\beta$-eucryptite calculated from the single crystal elastic constants predicted by
ReaxFF is in good agreement with those calculated using the elastic constants known
by DFT and experiments (refer to Table~\ref{table:polycrystalline}).

\section{Summary and conclusion}\label{sec:conclusion}

In conclusion, we have developed a reactive force field for lithium aluminum silicates
and used it to describe (i) the atomic structure and heats of formation of several oxides,
silicates and aluminates, (ii) the relative stability of three crystalline eucryptite polymorphs and the response of
$\beta$-eucryptite under indentation, and (iii) the anisotropic and polycrystalline-averaged elastic properties of eucryptite phases.

{\bf Successes.} We have found that structural properties and heats of formation for selected condensed phases agree well with
the results of DFT calculations and with experimental reports. In terms of applications to the stability of
eucryptite phases, we have verified that the order of the stability of three well-known polymorphs predicted by
ReaxFF is the same as that obtained from DFT calculations and that known from experiments. The response
of $\beta$-eucryptite to pressure is the formation of a denser and disordered phase which we
characterized by a set of radial distribution functions and comparisons with condensed phases.
In terms of elastic properties analysis, we have determined the elements of the stiffness tensor for $\alpha$- and $\beta$-
eucryptite at the level of ReaxFF, and discussed the elastic anisotropy of these two polymorphs.
Polycrystalline average properties of these eucryptite phases are also reported to serve as ReaxFF
predictions of their elastic moduli (in the case of $\alpha$-eucryptite), or as tests against values
known from experiments or DFT calculations ($\beta$-eucrypite).
    In addition to the elaborate but physically-motivated description of the bond order formalism coupled with the EEM
scheme, the novel aspects/results of this work include the ability of ReaxFF to predict the formation
of an amorphous phase under pressures exceeding 7 GPa, and the prediction of all elastic properties
of $\alpha$-eucryptite --which is the most stable LiAlSiO$_4$ phase at room temperature
and ambient pressure.

{\bf Shortcomings.} We noted in Sec.~IIIA that in the deformation regimes far outside equilibrium, the ReaxFF-predicted order of phase
stability may not match the DFT predictions, especially in the tensile regimes. This problem is likely to manifest during reaction
calculations at the level of several atoms, molecules, or small clusters, but may not easily manifest in
large-scale MD simulations because fracture in tensile regimes will probably occur before any phase transformation.
The values of the ReaxFF elastic constants $C_{ij}$ for $\beta$-eucryptite compare reasonably well
with those predicted by other empirical force fields.\cite{Beest1990,Winkler_THB1991,Cormack1996,Cormack1999,Pedone2006}
These values are not of superior accuracy, as they deviate by about 30\% from the experimental values.
Still, the values of $C_{ij}$ predicted by ReaxFF (Table~\ref{table:b-eucryptite}) deviate from experiments by amounts that
are very similar to the deviations made by the Pedone force field in predicting the elastic constants for spodumene (LiAlSi$_2$O$_6$).\cite{Pedone2006} However, we found that PFF\cite{Pedone2006} and a core-shell model potential developed
by Winkler {\em et al.} (THB)\cite{Winkler_THB1991} describe the elastic properties of aluminum silicate phases (especially those of andalusite,
Al$_2$SiO$_5$) much better than ReaxFF. While this observation seems to place ReaxFF at a disadvantage, it should be noted that the
PFF and other models were obtained by fitting against experimental values of the elastic properties of binary oxides and silicates,\cite{Pedone2006,Winkler_THB1991} while these properties were not a part of the training set used to determine the ReaxFF parameters in our study.

{\bf Concluding Remark.} The ReaxFF potential reported here can also describe well single-species systems ({\em e.g.}, Li-metal, Al-metal, and condensed phases of silicon), which makes it suitable for investigating structure and properties of suboxides, atomic-scale mechanisms responsible for phase transformations, as well as oxidation-reduction reactions. Based on the results of indentation on $\beta$-eucryptite and elastic properties of $\alpha$-eucryptite reported here, we believe that the parametrization of ReaxFF for Li-Al silicates will help provide fundamental understanding of
other interesting phenomena in LAS glass ceramics, especially in regard to the atomic scale mechanisms
underlying the pressure induced $\beta$-to-$\epsilon$ phase transformation where direct dynamic simulations at the level of DFT are
currently intractable.

{\em Acknowledgments.} The work at Colorado School of Mines was performed with support from the Department of Energy's Office
of Basic Energy Sciences through Grant No. DE-FG02-07ER46397 and from the National
Science Foundation (NSF) through Grant No. CMMI-0846858. ACTvD acknowledges funding from KISK startup grant
C000032472. We thank Prof. Jincheng Du from University of North Texas for providing the lithium-silicates DFT data that was published
in Ref.~\onlinecite{du}. Supercomputer time for the DFT calculations was provided by the Golden Energy Computing
Organization at Colorado School of Mines.

\newpage
\appendix
\setcounter{table}{0}
\renewcommand*\thetable{\Alph{section}.\Roman{table}}
\section{ReaxFF parameters for L\lowercase{i}-A\lowercase{l}-S\lowercase{i}-O systems}\label{appendix:param}
\setcounter{table}{0}
The ReaxFF parameters for the Li-Al-Si-O systems determined in the present study are listed in Tables~\ref{table:general_parameters}$-$\ref{table:valence_angle_parameters}.

\begin{table}[h]
\caption{General Parameters} \label{table:general_parameters}
\begin{center}
\begin{tabular*}{0.45\textwidth}{@{\extracolsep{\fill}}lcc}
\toprule
%&&&&&&\\
 Parameter               &  Value     & Description \\
\colrule
%                   &&           &&
$p_{boc1}$          &  50.0000   & Bond order correction            \\
%                   &&           &&
$p_{boc2}$          &   9.5469   & Bond order correction            \\
%             	    &&           &&	
$p_3$               &  50.0000   & Overcoordination             \\
%                   &&           &&
$p_4$               &   0.6991   & Overcoordination             \\
%                   &&           &&
$p_{6}$             &   1.0588   & Undercoordination                                 \\
%                   &&           &&
$p_{7}$             &  12.1176   & Undercoordination             	\\
%                   &&           &&
$p_{8}$             &  13.3056   & Undercoordination             \\
%                   &&           &&
$p_{lp1}$           &   6.0891   & Lone pair parameter            \\
%                   &&           &&
$p_{v7}$            &  33.8667   & Valence undercoordination            \\
%                   &&           &&
$p_{v8}$            &   1.8512   & Valence angle            \\
%                   &&           &&
$p_{v9}$            &   1.0563   & Valence angle             \\
%                   &&           &&
$p_{v10}$           &   2.0384   & Valence angle            \\
%                   &&           &&
$p_{vdW1}$          &   1.5591   & van der Waals shielding            \\
%                   &&           &&
$BO_{cut}$          &   0.0010   & Bond order cut-off            \\
\botrule
\end{tabular*}
\end{center}
\end{table}

\begin{table}[h]
\caption{Atom parameters. All the parameters except $p_{lp2}$ (kcal/mol) are unitless} \label{table:atom_parameters}
\begin{center}
\begin{tabular*}{0.45\textwidth}{@{\extracolsep{\fill}}lcccccc}
\toprule
%&&&&&&\\
 Atom               &  $\mathcal{V}_i$ & $\mathcal{V}_i^e$ & $\mathcal{V}_i^a$ & $\mathcal{V}_i^{boc}$ & $p_{2}$ & $p_{5}$\\
\colrule
%                   &&            &&         &&                 \\
Li                  & 1.0000 & 1.0000 & 1.0000 & 1.0000 & -24.7916  & 0.0000 \\
%                   &&            &&         &&                 \\
Al                  & 3.0000 & 3.0000 & 3.0000 & 8.0000 & -23.1826  & 0.0076 \\
%             	    &&            &&         &&                 \\
Si                  & 4.0000 & 4.0000 & 4.0000 & 4.0000 & -4.1684   & 21.7115 \\
%                   &&            &&         &&                 \\
O                   & 2.0000 & 6.0000 & 4.0000 & 4.0000 & -3.5500   & 37.5000 \\
                    &        &        &        &        &           &          \\
\colrule
                    & $p_{v3}$ & $p_{v5}$ & $p_{lp2}$ & $p_{boc3}$ &  $p_{boc4}$ & $p_{boc5}$       \\
\colrule		
Li		    & 2.2989   & 2.8103   & 0.0000    & 6.9107     & 5.4409      & 0.1973           \\
Al		    & 1.5000   & 2.5791   & 0.0000    & 0.2500     & 20.0000     & 0.0000           \\
Si		    & 2.0754   & 2.5791   & 0.0000    & 23.8188    & 9.0751      & 0.8381           \\
O		    & 2.9000   & 2.9225   & 0.4056    & 0.7640     & 3.5027      & 0.0021           \\
\botrule
\end{tabular*}
\end{center}
\end{table}

\begin{table}
\caption{Covalent radii [$r_0^\sigma$, $r_0^\pi$, $r_0^{\pi\pi}$ in \AA] and Coulomb interaction parameters [$\eta$ (eV), $\chi$ (eV) and $\gamma$ (\AA)].} \label{table:coulomb_parameters}
\begin{center}
\begin{tabular*}{0.48\textwidth}{@{\extracolsep{\fill}}lcccccc}
\toprule
%&&&&&&\\
                    &                     &           &                &\multicolumn{3}{c}{Coulomb parameters}\\
\cline{5-7}
Atom                &  $r_0^\sigma$  & $r_0^\pi$ & $r_0^{\pi\pi}$ & $\eta$      & $\chi$      & $\gamma$ \\
%                    &  (\AA)         &    (\AA)  &    (\AA)       &   (eV)      & (eV)        &    (\AA)      \\
\colrule
%                   &&            &&         &&                 \\
Li                  &  1.6908   & -0.1000  & -1.0000 & 11.0234 & -3.2182 & 1.0000       \\
%                   &&            &&         &&                 \\
Al                  &  2.1967   & -1.6836  & -1.0000 & 6.5000  & -0.3343 & 0.4961  \\
%             	    &&            &&         &&                 \\
Si                  &  2.1932   &  1.2962  & -1.0000 & 5.5558  & 4.2033  & 0.5947  \\
%                   &&            &&         &&                 \\
O                   &  1.2450   &  1.0548  &  0.9049 & 8.3122  & 8.5000  & 1.0898  \\
\botrule
\end{tabular*}
\end{center}
\end{table}

\begin{table}
\caption{Van der Waals interaction parameters.} \label{table:vdw_parameters}
\begin{center}
\begin{tabular*}{0.45\textwidth}{@{\extracolsep{\fill}}lcccc}
\toprule
%&&&&&&\\
Atom                &  $r_{vdW}$ (\AA) & $D_{ij}$ (kcal/mol) & $\alpha$ & $\gamma_{vdW}$ (\AA)\\
\colrule
%                   &&            &&         &&                 \\
Li                  &  1.6121   & 0.2459 & 10.8333  & 1.4649       \\
%                   &&            &&         &&                 \\
Al                  &  2.3738   & 0.2328 &  9.4002  & 1.6831  \\
%             	    &&            &&         &&                 \\
Si                  &  1.8951   & 0.1737 & 11.3429  & 5.2054  \\
%                   &&            &&         &&                 \\
O                   &  2.3890   & 0.1000 &  9.7300  & 13.8449  \\
\botrule
\end{tabular*}
\end{center}
\end{table}

\begin{table}
\caption{Bond parameters. The bond dissociation energies $D_e^\sigma$, $D_e^\pi$ and $D_e^{\pi\pi}$ are in kcal/mol while $p_{be1}$, $p_{be2}$ and $p_1$ are unitless} \label{table:be_parameters}
\begin{center}
\begin{tabular*}{0.5\textwidth}{@{\extracolsep{\fill}}lcccccc}
\toprule
%&&&&&&\\
        & $D_e^\sigma$	& $D_e^{\pi}$	& $D_e^{\pi\pi}$& $p_{be1}$	& $p_{be2}$	  & $p_{1}$ \\
%        & (kcal/mol)	& (kcal/mol)	& (kcal/mol)	&		&		  &         \\
\colrule	
O$-$O	& 142.2858 	& 145.0000 	& 50.8293	&  0.2506	& -0.1055	  & 0.3451  \\
Si$-$O	& 274.8339	&  5.0000	&  0.0000	& -0.5884	& -0.2572	  & 9.9772  \\
Si$-$Si	&  70.9120	& 54.0531	& 30.0000	&  0.4931	& -0.8055	  & 0.2476  \\
Al$-$O	& 181.1998	&  0.0000	&  0.0000	& -0.2276	& -0.3500	  & 0.2086  \\
Al$-$Si	&   0.0000	&  0.0000	&  0.0000	&  1.0000	&  0.0000	  & 0.5000  \\
Al$-$Al	&  34.0777	&  0.0000	&  0.0000	&  0.4832	& -0.4197	  & 6.4631  \\
Li$-$O	&  78.3666	& -0.0200	&  0.0000	& -1.0000	& -0.2500	  & 0.2022  \\
Li$-$Si	&   0.0000	&  0.0000	&  0.0000	&  1.0000	&  0.0000	  & 0.5000  \\
Al$-$Li	&   0.0000	&  0.0000	&  0.0000	&  1.0000	&  0.0000	  & 0.5000  \\
Li$-$Li	&  42.9780	&  0.0000	&  0.0000	&  0.3228	&  0.0000	  & 1.7161  \\
\botrule
\end{tabular*}
\end{center}
\end{table}

\begin{table}
\caption{Bond order parameters.} \label{table:bo_parameters}
\begin{center}
\begin{tabular*}{0.5\textwidth}{@{\extracolsep{\fill}}lcccccc}
\toprule
%&&&&&&\\
Bond                &  $p_{bo,1}$ & $p_{bo,2}$ & $p_{bo,3}$ & $p_{bo,4}$ & $p_{bo,5}$ & $p_{bo,6}$\\
\colrule
O$-$O		    & 5.5000	& 1.0000	&  9.0000	& 1.0000 	& -0.1000	& 0.6051  \\
Si$-$O		    & 8.4790	& 6.0658	& 28.8153	& 1.0000	& -0.3000	& 0.2131  \\
Si$-$Si		    & 8.7229	& 0.0000	&  7.1248	& 1.0000	& -0.3000	& 0.0392  \\
Al$-$O		    & 6.1462	& 0.0000	& 25.0000	& 1.0000	& -0.3000	& 0.1925  \\
Al$-$Si		    & 10.0000	& 0.0000	& 12.0000	& 1.0000	&  0.3000	& 1.0000  \\
Al$-$Al		    & 6.1608	& 0.0000	& 14.3085	& 1.0000	& -0.3000	& 0.5154  \\
Li$-$O		    & 7.8656	& 0.0000	& 11.9965	& 1.0000	&  0.3000	& 0.3228  \\
Li$-$Si		    & 10.0000	& 0.0000	& 12.0000	& 1.0000	&  0.3000	& 1.0000  \\
Al$-$Li		    & 10.0000	& 0.0000	& 12.0000	& 1.0000	&  0.3000	& 1.0000  \\
Li$-$Li		    & 4.0000	& 0.0000	& 12.0000	& 1.0000	&  0.3000	& 0.6003  \\
\botrule
\end{tabular*}
\end{center}
\end{table}

\begin{table}
\caption{Off-diagonal bond parameters [$D_{ij}$ (kcal/mol), $\alpha$ (unitless)] and bond radii [$R_{vdW}$, $r_0^{\sigma}$, $r_0^{\pi}$, and $r_0^{\pi\pi}$ (\AA)].} \label{table:off_diagonal_bond_parameters}
\begin{center}
\begin{tabular*}{0.5\textwidth}{@{\extracolsep{\fill}}lcccccc}
\toprule
%&&&&&&\\
Bond                &  $D_{ij}$ &  $R_{vdW}$ & $\alpha$ & $r_0^{\sigma}$ & $r_0^{\pi}$ & $r_0^{\pi\pi}$\\
\colrule
%                   &&            &&         &&                 \\
Si$-$O              &  0.1836   & 1.9157 & 10.9070 & 1.7073  & 1.2375  & -1.0000  \\
%                   &&            &&         &&                                \\
Al$-$O              &  0.2017   & 1.8458 & 11.0700 & 1.6009  & -1.0000 & -1.0000 \\
%                   &&            &&         &&                 \\
Al$-$Si             &  0.1000   & 1.8500 & 10.3237 & -1.0000 & -1.0000 & -1.0000 \\
%                   &&            &&         &&                 \\
Li$-$O              &  0.0790   & 2.2000 & 9.0491  & 1.8165  & -1.0000 & 1.0000  \\
%                   &&            &&         &&                 \\
Li$-$Si             &  0.0200   & 1.5000 & 10.0529 & -1.0000 & 1.0000  & 1.0000 \\
%                   &&            &&         &&                 \\
Li$-$Al             &  0.1146   & 2.2000 & 9.7537  & -1.0000 & 1.0000  & 1.0000 \\
\botrule
\end{tabular*}
\end{center}
\end{table}

\begin{table}
\caption{Valence angle parameters} \label{table:valence_angle_parameters}
\begin{center}
\begin{tabular*}{0.47\textwidth}{@{\extracolsep{\fill}}lccccc}
\toprule
%&&&&&&\\
                 &  ${\Theta}_{0,0}$ & $p_{v1}$ & $p_{v2}$ & $p_{v4}$ & $p_{v7}$ \\
		 &  (deg.)           & (kcal/mol) &        &          &          \\
\colrule
O$-$O$-$O	         & 80.7324	& 30.4554	& 0.9953 	& 1.0783	& 1.6310 \\
Si$-$Si$-$Si         	 & 78.5339	& 36.4328	& 1.0067 	& 1.6608	& 0.1694 \\
O$-$Si$-$Si	         & 86.3294	& 18.3879	& 5.8529 	& 1.2310	& 1.7361 \\
O$-$Si$-$O	 	 & 79.5581	& 34.9140	& 1.0801	& 2.2206	& 0.1632 \\
Si$-$O$-$Si	 	 & 82.3364	&  4.7350	& 1.3544	& 1.0400	& 1.4627 \\
O$-$O$-$Si	 	 & 92.1207	& 24.3937	& 0.5000	& 3.0000	& 1.7208 \\
O$-$O$-$Al	 	 & 34.4326	& 25.9544	& 5.1239	& 1.7141	& 2.7500 \\
Al$-$O$-$Al	 	 & 20.7204	& 13.4875	& 4.0000	& 1.4098	& 0.6619 \\
O$-$Al$-$O	 	 & 59.5433	& 20.0000	& 4.0000	& 2.0988	& 3.0000 \\
O$-$Li$-$O		 & 60.0000	&  0.0000	& 1.0000	& 1.0000	& 1.0000 \\
O$-$O$-$Li		 & 81.6233	& 30.0000	& 2.0000	& 1.0000	& 1.0000 \\
Li$-$O$-$Li		 & 67.5247	&  6.4512	& 4.0000	& 2.8079	& 1.0000 \\
Al$-$O$-$Li		 & 50.9423	&  7.0901	& 3.9271	& 2.5544	& 1.0000 \\
Si$-$O$-$Al		 & 18.0953	&  5.3220	& 4.0000	& 1.0139	& 1.0000 \\
Si$-$O$-$Li		 & 62.6634	&  8.4441	& 2.5120	& 1.0000	& 1.0000 \\
\botrule
\end{tabular*}
\end{center}
\end{table}\newpage

%\begin{table}[htbp]
%\caption{Torsion angle parameters, $V_1$, $V_2$, and $V_3$ (kcal/mol).} \label{table:torsion_angle_parameters}
%\begin{center}
%\begin{tabular*}{0.5\textwidth}{@{\extracolsep{\fill}}lccccc}
%\toprule
%&&&&&&\\
%Torsion Angle       &  $V_1$ &  $V_2$ & $V_3$ & $p_t$ & $p_{conj}$\\
%\colrule
%                   &&            &&         &&                 \\
%O$-$O$-$O$-$O       &  -2.5  & -4.0   & 1.0   & -2.5 & -1.0   \\
%                   &&            &&         &&                                \\
%\botrule
%\end{tabular*}
%\end{center}
%\end{table}\newpage

\setcounter{table}{0}
\section{Calculation of elastic constants} \label{appendix:elastic_calc}

The elements of the elastic stiffness tensor $C_{ijkl}$ for $\alpha$ and $\beta$ eucryptite
were computed within the framework of ReaxFF by calculating the second derivatives of strain energy density
with respect to the strain components\cite{Wallace}
\begin{equation}
C_{ijkl} = \frac{\partial^2(E/V)}{\partial\epsilon_{ij}\epsilon_{kl}},
\end{equation}
where $E$ is the elastic energy stored in a domain of volume $V$ of the crystal subjected to homogeneous deformations.
A similar approach has been employed earlier for computing the elastic constants of $\beta$-eucryptite using
DFT  calculations (See Ref.~\onlinecite{Narayanan2010}). For sufficiently small strains, the total energy $E$ of a
crystal subjected to a general strain can be expressed as a Taylor series expansion truncated
at the second order\cite{Wallace}
\begin{equation} \label{Wallace_eq}
E(V,\mathbf{\epsilon})=E_0+V_0\left(\sum_{i}\sigma_{i}\epsilon_{i}\eta_{i}+\sum_{i,j}\frac{1}{2}C_{ij}\epsilon_{i}\eta_{i}\epsilon_{j}\eta_{j}\right),
\end{equation}
where the subscripts are cast in the Voigt notation (11=1, 22=2, 33=3, 23=4, 31=5, and 12=6),
$\eta_i =1 $ if  $i = 1, 2, \mbox{or } 3$ and $\eta_i= 2$ if $i = 4, 5, \mbox{or } 6$, $E_0$ is the energy of the
crystal volume $V_0$ at equilibrium, ${\sigma}_{ij}$ are the elements of the stress tensor, and $\delta_{ij}$ is the Kronecker symbol.
For the strains listed in Tables~\ref{table:hexagonal_elastic} and~\ref{table:rhombohedral_elastic}, Eq.~(\ref{Wallace_eq}) reduces
to
\begin{equation} \label{eq:final}
E(V,\delta) = E_0 + V_0(A_1\delta + A_2{\delta}^{2}),
\end{equation}
where $A_1$ is related to stress components $\sigma_{ij}$, and $A_2$ is a linear combination of the elastic
constants $C_{ij}$ expressed in the Voigt notation.

$\beta$-eucryptite has five \emph{independent} elastic constants namely, $C_{11}$, $C_{12}$, $C_{13}$, $C_{33}$ and $C_{44}$ due to the hexagonal symmetry associated with its structure.~\cite{Nye} Table~\ref{table:hexagonal_elastic} lists the five different strains that we utilised to compute the elastic constants of $\beta$-eucryptite along with the relationship between the second-order coefficient $A_2$ and the elastic constants $C_{ij}$ for each type of strain.
\begin{table}
\caption{The strains used to calculate the five independent elastic constants of
hexagonal $\beta$-eucryptite (also used in Refs.~\onlinecite{Fast1995} and~\onlinecite{Narayanan2010}).
The relationship between $A_2$ in Eq.~(\ref{eq:final}) and $C_{ij}$ are also provided.}\label{table:hexagonal_elastic}
\begin{center}
\begin{tabular}{cc}
\toprule
%&&&&&&&\\
 Strain parameters       &  Second-order coefficient \\
(unlisted $\epsilon_i = 0$) & $A_2$ in Eq.~(\ref{eq:final})       \\

\colrule
%            &           &        &            \\
$\epsilon_1 = \epsilon_2 = \delta$         &  $C_{11}+C_{12}$   \\
%            &           &        &            \\
$\epsilon_1 = -\epsilon_2 = \delta$        &  $C_{11}-C_{12}$   \\
%            &           &        &            \\
$\epsilon_1 = \epsilon_2 = \epsilon_3 = \delta$     &  $C_{11} + C_{12} + 2C_{13} + C_{33}/2$   \\
%            &           &        &            \\
$\epsilon_3 = \delta$                                &  $C_{33}/2$     \\
%            &           &        &            \\
$\epsilon_5 = \delta$                                &  $2C_{44}$ \\
%            &           &        &            \\

\botrule
\end{tabular}
\end{center}
%\caption{Comparison of stiffness constants ${C}_{ij}$  of andalusite with experimental data.}
\end{table}

\begin{table}
\caption{The strains used to calculate the seven independent elastic constants of rhombohedral $\alpha$-eucryptite.
The relationship between $A_2$ in Eq.~(\ref{eq:final}) and $C_{ij}$ are also provided.}\label{table:rhombohedral_elastic}
\begin{center}
\begin{tabular}{ccc}
\toprule
%&&&&&&&\\
Strain parameters       &&  Second-order coefficient \\
(unlisted $\epsilon_i = 0$) && $A_2$ in Eq.~(\ref{eq:final})       \\
\colrule
%            &           &        &            \\
$\epsilon_1 = \delta$     &&  $C_{11}/2$   \\
%            &           &        &            \\
$\epsilon_3 = \delta$     &&  $C_{33}/2$   \\
%            &           &        &            \\
$\epsilon_1 = \epsilon_2 = \epsilon_3 = \delta$    && $C_{11} + C_{12} + 2C_{13} + C_{33}/2$      \\
%            &           &        &            \\
$\epsilon_1 = -\epsilon_2 = \epsilon_3 = \delta$    && $C_{11} - C_{12} + C_{33}/2$      \\
%            &           &        &            \\
$\epsilon_1 = \epsilon_4 = \delta$       && $C_{11}/2 + 2C_{14} + 2C_{44}$      \\
%            &           &        &            \\
$\epsilon_1 =\epsilon_5 = \delta$       && $C_{11}/2 + 2C_{15} + 2C_{44}$      \\
%            &           &        &            \\
$\epsilon_4 = \delta$     &&  $2C_{44}$   \\

%            &           &        &            \\
\botrule
\end{tabular}
\end{center}
\end{table}

On the other hand, $\alpha$-eucryptite has a rhombohedral crystal structure and thereby,
has seven \emph{independent} elastic constants namely, $C_{11}$, $C_{12}$, $C_{13}$, $C_{14}$, $C_{15}$, $C_{33}$ and $C_{44}$.\cite{Nye}
The different strains used to compute these seven elastic constants and the relationships
between $A_2$ and $C_{ij}$ for each type of strain have been summarized in Table~\ref{table:rhombohedral_elastic}.

For a given crystal, the total energy was computed for different values of $\delta$ ranging
from -2\% to 2\% using the LAMMPS~\cite{LAMMPS} implementation of ReaxFF. The calculated data were
then fit to Eq.~(\ref{eq:final}) to extract the second-order coefficients $A_2$ which were then
used to evaluate the elastic constants through the relationships given in
Tables~\ref{table:hexagonal_elastic} and~\ref{table:rhombohedral_elastic}.\newpage

%\bibliography{biblio/ref.bib}

\begin{thebibliography}{99}

\bibitem{Bach} H. Bach, ed., \emph{Low Thermal Expansion Glass Ceramics}, Schott Series on Glass and Glass ceramics (Springer, Berlin, 1995).

\bibitem{Palmer} D.~C. Palmer, in \emph{Reviews in Mineralogy}, edited by P.~J. Heaney, C.~T. Prewitt, and G.~V. Gibbs (Mineralogical Society of America, Washington D.~C.,1996), vol. 29, p. 83.

\bibitem{Xu1999} H. Xu, P.~J. Heaney, D.~M. Yates, R.~B. von Dreele, and M.~A. Bourke, J. Mater. Res. \textbf{14}, 3138 (1999).

\bibitem{Lichtenstein1998} A.~I. Lichtenstein, R.~O. Jones, H. Xu, P.~J. Heaney, Phys. Rev. B \textbf{58}, 6219 (1998).

\bibitem{Lichtenstein2000} A.~I. Lichtenstein, R.~O. Jones, S. de Gironcoli, and S. Baroni, Phys. Rev. B \textbf{62}, 11487 (2000).

\bibitem{Winkler} H.~G.~F. Winkler, Acta Crystallogr. \textbf{1}, 27 (1948).

\bibitem{Buerger} M.~J. Buerger, Am. Mineral. \textbf{39}, 600 (1954).

\bibitem{Schulz_B28_1972_II} H. Schulz and V. Tscherry, Acta Crystallogr., Sect. B: Struct. Crystallogr. Cryst. Chem. \textbf{28}, 2174 (1972).

\bibitem{Tscherry_161_1972} V. Tscherry, H. Schulz, and F. Laves, Z. Kristallogr. \textbf{135}, 161 (1972).

\bibitem{Tscherry_175_1972} V. Tscherry, H. Schulz, and F. Laves, Z. Kristallogr. \textbf{135}, 175 (1972).

\bibitem{Pillars1973} W.~W. Pillars and D.~R. Peacor, Am. Mineral. \textbf{58}, 681 (1973).

\bibitem{Alpen1977} U.~V. Alpen, H. Schulz, G.~H. Talat, and H. B{\"o}hm, Sol. Stat. Comm. \textbf{23}, 911 (1977).

\bibitem{Schulz1980} W. Press, B. Renker, H. Schulz, and H. B{\"o}hm, Phys. Rev. B \textbf{21}, 1250 (1980).

\bibitem{Nagel1982} W. Nagel and H. B{\"o}hm, Sol. Stat. Comm. \textbf{42}, 625 (1982).

\bibitem{Renker1983} B. Renker, H. Bernotat, G. Heger, N. Lehner, and W. Press, Sol. Stat. Ion. \textbf{9}, 1341 (1983).

\bibitem{Sartbaeva2004} A. Sartbaeva, S.~A. Wells, and S.~A.~T.Redfern, J. Phys. Cond. Mat. \textbf{16}, 8173 (2004).

%\bibitem{Richet1997} P. Richet and P.Gillet, Eur. J. Mineral. \textbf{9}, 907 (1997).

\bibitem{Zhang2002} J. Zhang, A. Celestian, J.~B. Praise, H. Xu, and P.~J. Heaney, Am. Mineral. \textbf{87}, 566 (2002).

\bibitem{Jochum2009} T. Jochum, I.~E. Reimanis, M.~J. Lance and E.~R. Fuller, Jr., J. Am. Ceram. Soc. \textbf{92}, 857 (2009).

\bibitem{Beall} G.~H. Beall, in \emph{Silica: Physical behavior, Geochemistry and Physical Applications}, edited by P.~J. Heaney, C.~T. Prewitt, and G.~V. Gibbs (Mineralogical Society of America, 1994), p. 469.

\bibitem{Zhang2005} J. Zhang, Y. Zhao, H. Xu, M.~V. Zelinskas, L. Wang, Y. Wang, and T. Uchida, Chem. Mater. \textbf{17}, 2817 (2005).

\bibitem{Daniels2001} P. Daniels and C.~A. Fyfe, Am. Mineral. \textbf{86}, 279 (2001).

\bibitem{Fasshauer1998} D.~W. Fasshauer, N.~D. Chatterjee, and L. Cemic, Contrib. Mineral. Petrol. \textbf{133}, 186 (1998).

\bibitem{m1} R. Marto\v{n}\'{a}k, A. Laio, and M. Parrinello, Phys. Rev. Lett.  {\bf 90}, 075503 (2003).

\bibitem{m2} D. Donadio, R. Marto\v{n}\'{a}k, P. Raiteri, and M. Parrinello, Phys. Rev. Lett. {\bf 100}, 165502 (2008).

\bibitem{Lewis1985} G.~V. Lewis and C.~R.~A. Catlow, J. Phys. C \textbf{18}, 1149 (1985).

\bibitem{Huang1990} C. Huang, Ph.D. thesis, Alfred University (1990).

\bibitem{Beest1990} B.~W.~H. van Beest, G.~J. Kramer, and R.~A. van Santen, Phys. Rev. Lett. \textbf{64}, 1955 (1990).

\bibitem{Winkler_THB1991} B. Winkler, M.~T. Dove, and M. Leslie, Am. Mineral. \textbf{76}, 313 (1991).

\bibitem{Cormack1996} A.~N. Cormack and Y. Cao, Mol. Engg. \textbf{6}, 183 (1996).

\bibitem{Cormack1999} A.~N. Cormack, X. Yuan, and B. Park, Glass Phys. Chem. \textbf{27}, 28 (1999).

\bibitem{Pedone2006} A. Pedone, G. Malavski, M.~C. Menziani, A.~N. Cormack, and U. Segre, J. Phys. Chem. \textbf{110} (2006).

\bibitem{VanDuin2001} A.~C.~T. van Duin, S. Dasgupta, F. Lorant, and W.~A. Goddard, III, J. Phys. Chem. A \textbf{105}, 9396 (2001).

\bibitem{VanDuin2003} A.~C.~T. van Duin, A. Strachan, S. Stewman, Q. Zhang, X. Xu, and W.~A. Goddard, III, J. Phys. Chem. A \textbf{107}, 3803 (2003).

\bibitem{Tersoff1988} J. Tersoff, Phys. Rev. Lett. \textbf{61}, 2879 (1988).

\bibitem{Benner1990} D.~W. Benner, Phys. Rev. B \textbf{42}, 9458 (1990).

\bibitem{Rappe1991} A. Rapp{\'e} and W.~A. Goddard, III, J. Phys. Chem. \textbf{95}, 3358 (1991).

\bibitem{LAMMPS} S. Plimpton, J. Comp. Phys. \textbf{117}, 1 (1995).

\bibitem{Strachan2005} A. Strachan, E.~M. Kober, A.~C.~T. van Duin, J. Oxgaard, and W.~A. Goddard, III, J. Chem. Phys. \textbf{122}, 054502 (2005).

\bibitem{Buehler2006} M. Buehler, A.~C.~T. van Duin, and W.~A. Goddard, III, Phys. Rev. Lett. \textbf{96}, 095505 (2006).

\bibitem{Zhang2004} Q. Zhang, T. {\c{C}}a{\u{g}}in, A.~C.~T van Duin, W.~A. Goddard, III, Y. Qi, and L.~G. Hector, Jr., Phys. Rev. B \textbf{69}, 045423 (2004).

\bibitem{Raymand2008} D. Raymand, A.~C.~T. van Duin, M. Baudin, and K. Hermansson, Surf. Sci. \textbf{602}, 1020 (2008).

\bibitem{VanDuin2008} A.~C.~T. van Duin, B.~V. Merinov, S.~S. Jang, and W.~A. Goddard, III, J. Phys. Chem. A \textbf{112}, 3133 (2008).

\bibitem{Goddard2002} W.~A. Goddard, III, Q. Zhang, M. Uludogan, A. Strachan, and T. {\c{C}}a{\u{g}}in, in \emph{Fundamental Physics of
    Ferroelectrics}, edited by R.~E. Cohen (American Institute of Physics, 2002), p. 45.

\bibitem{Pauling1947} L. Pauling, J. Am. Chem. Soc. \textbf{69}, 542 (1947).

\bibitem{vanDuin_book} A. van Duin, in \emph{Computational Methods in Catalysts and Materials Science}, edited by R. van Santen and P. Sautet (Wiley-VCH Verlag Gmbh \& Co., Weinheim, Germany, 2009), p. 167.

\bibitem{Han2005} S.~S. Han, A.~C.~T. van Duin, W.~A. Goddard, III, and H.~M. Lee, J. Phys. Chem. A \textbf{109}, 4575 (2005).

\bibitem{Ojwang2008} J.~G.~O. Ojwang, R. van Santen, G. J. Kramer, A.~C.~T. van Duin, and W.~A. Goddard, III, J. Chem. Phys. \textbf{128}, 164714 (2008).

\bibitem{Ojwang2009} J.~G.~O. Ojwang, R. van Santen, G.~J. Kramer, A.~C.~T. van Duin, and W.~A. Goddard, III, J. Chem. Phys \textbf{131}, 044501 (2009).

\bibitem{Cheung2005} S. Cheung, W.~Q. Deng, A.~C.~T. van Duin, and W.~A. Goddard, III, J. Phys. Chem. A \textbf{109}, 851 (2005).

\bibitem{du} J. Du and L.~R. Corrales, J. Phys. Chem. B \textbf{110}, 22346 (2006).

\bibitem{Lazicki2006} A. Lazicki, C.~S. Yoo, W.~J. Evans, and W.~E. Pickett, Phys. Rev B \textbf{73}, 184120 (2006).

\bibitem{Cota2005} L.~G. Cota and P. de la Lora, Acta Crystallogr., Sect. B: Struct. Sci. \textbf{61}, 133 (2005).

\bibitem{Marezio_alpha1965} M. Marezio and J.~P. Remeika, J. Chem. Phys. \textbf{44}, 3143 (1966).

\bibitem{Dronskowski1993} R. Dronskowski, Inorg. Chem. \textbf{32}, 1 (1993)

\bibitem{Wu2009} S.~Q. Wu, Z.~F. Hou, and Z.~Z. Zhu, Comp. Mat. Sci. \textbf{46}, 221 (2009).

\bibitem{Winter1979} J.~K. Winter and S. Ghose, Am. Mineral. \textbf{64}, 573 (1979).

\bibitem{Ralph1984} R.~L. Ralph, L.~W. Finger, R.~M. Hazen, and S. Ghose, Am. Mineral. \textbf{69}, 513 (1984).

\bibitem{Yang_PCM1997} H. Yang, R.~M. Hazen, L.~W. Finger, C.~T. Prewitt, and R.~T. Downs, Phys. Chem. Mineral. \textbf{25}, 39 (1997).

\bibitem{Yang_AmM1997} H. Yang, R.~T. Downs, L.~W. Finger, R.~M. Hazen, and C.~T. Prewitt, Am. Mineral. \textbf{82}, 467 (1997).

\bibitem{VanDuin1994} A.~C.~T. van Duin, J.~M.~A. Baas, and B. van de Graaf, J. Chem. Soc., Farad. Trans. \textbf{90}, 2881 (1994).

\bibitem{Text2010} See EPAPS Document No. [to be given by the publisher] for ReaxFF parameters in a format compatible with LAMMPS.

\bibitem{PAW} G. Kresse and J. Joubert, Phys. Rev. B \textbf{59}, 1758 (1996).

\bibitem{VASP1} G. Kresse and J. Furthm{\"u}ller, Comp. Mat. Sci. \textbf{6}, 15 (1996).

\bibitem{VASP2} G. Kresse and J. Furthm{\"u}ller, Phys. Rev. B \textbf{54}, 11169 (1996).

\bibitem{PW91} J.~P. Perdew, J.~A. Chevary, S.~H. Vosko, K.~A. Johnson, M.~R. Pederson, D.~J. Singh, and C. Fiolhais, Phys. Rev. B \textbf{46}, 6671
    (1992).

\bibitem{Chase1988} M.~W. Chase, Jr., ed., \emph{NIST-JANAF Thermochemical tables Part I and II.} (Springer, Berlin, Heidelberg, New York, 1988), 4th ed., Journal of Physical and Chemical Reference Data.

\bibitem{Robie1995} R.~A. Robie and B.~S. Hemingway, U. S. Geological Survey Bulletin \textbf{2131}, 461 (1995).

\bibitem{Stolen2004} S. St{\o}len, T. Grande, and N.~L. Allan, \emph{Chemical thermodynamics of materials: Macroscopic and Microscopic aspects} (John Wiley and Sons Limited, West Sussex, England, 2004), p. 11.

\bibitem{Nakagawa1981} H. Nakagawa, M. Asano, and K. Kubo, J. Nucl. Mater. \textbf{102}, 292 (1981).

\bibitem{Heaney1999} H. Xu, P.~J. Heaney, A. Navrotsky, L. Topor, and J. Liu, Am. Mineral. \textbf{84}, 1360 (1999).

\bibitem{Kunc2005} K. Kunc, I. Loa, A. Grzechnik, and K. Syassen, Phys. Stat. Solidi. B \textbf{242}, 1857 (2005).

\bibitem{Marezio1965} M. Marezio, Acta Crystallogr. \textbf{19}, 396 (1965).

\bibitem{Winkler2001} B. Winkler, M. Hytha, M.~C. Warren, V. Milman, J.~D. Gale, and J. Shreuer, Z. Kristallogr. \textbf{216}, 67 (2001).

\bibitem{Tang2010} T. Tang and D.~J. Luo, J. At. Mol. Sci. \textbf{1}, 185 (2010).

\bibitem{Hesse1977} K.~F. Hesse, Acta Crystallogr., Sect. B: Struct. Crystal. Chem. \textbf{33}, 901 (1977).

\bibitem{Narayanan2010} B. Narayanan, I.~E. Reimanis, E.~R. Fuller, and C.~V. Ciobanu, Phys. Rev. B \textbf{81}, 104106 (2010).

\bibitem{Zhang1993} X. Zhang and C.~K. Ong, Phys. Rev. B \textbf{48}, 6865 (1993).

\bibitem{Rino2004} J.~P. Rino, I.Ebbsj{\"o}, P.~S. Branicio, R.~K. Kalia, A. Nakano, F. Shimojo and P. Vashishta, Phys. Rev. B \textbf{70}, 045207 (2004).

\bibitem{Watson1995} G.~W. Watson and S.~C. Parker, Phys. Rev. B \textbf{52}, 13306 (1995).

\bibitem{Bellert2001} D. Bellert and W.~H. Breckenridge, J. Chem. Phys. \textbf{114}, 2871 (2001).

\bibitem{Weast1987} R.~C. Weast, ed., \emph{Handbook of Chemistry and Physics}, 68th Ed (CRC Press, Boca Raton FL, 1987).

\bibitem{Born1956} M. Born, and K. Huang, \emph{Dynamical Theory of Crystal Lattices} (Clarendon, Oxford, 1956).

\bibitem{Haussuhl1984} S. Hauss{\"u}hl, W. Nagel, and H. B{\"o}hm, Z. Kristallogr. \textbf{169}, 299 (1984).

\bibitem{Nye} J.~F. Nye,\emph{Physical Properties of Crystals} (Oxford University Press, Oxford, 1985).

\bibitem{Voigt1928} W. Voigt, \emph{Lehrbook der kristallphysik} (Teubner, Leipsig, 1928).

\bibitem{Reuss1929} A.~Z. Reuss, Angew. Math. Mech. \textbf{9}, 49 (1929).

\bibitem{Hill1952} R. Hill, Proc. Phys. Soc. London Sect. A \textbf{65}, 349 (1952).

\bibitem{Simmons1971} G. Simmons and H. Wang, \emph{Single Crystal Elastic Constants and Calculated Aggregate Properties} (MIT, Cambridge, Massachusetts, 1971), 2nd ed.

\bibitem{Wallace} D.~C. Wallace, \emph{Thermodynamics of crystals} (Wiley, New York, 1972).

\bibitem{Fast1995} L. Fast, J.~M. Wills, B. Johansson, and O. Eriksson, Phys. Rev. B \textbf{51}, 17431 (1995).

\end{thebibliography}

%\begin{comment}

\end{document}